\documentclass[preprint,twoside]{elsarticle} 

\usepackage[T1]{fontenc}
\usepackage[cmex10]{amsmath}
\usepackage{amssymb,amsthm}
\usepackage{dsfont}
\usepackage{mathtools}
\usepackage{bm}
\usepackage{url}
\usepackage{booktabs}
\usepackage{tabularx}
\usepackage{tikz}
\usepackage{enumitem}
\usepackage{xcolor}
\usepackage{sistyle}
\usepackage[title]{appendix}
\SIthousandsep{,}

\usepackage[breaklinks,colorlinks,citecolor=blue,linkcolor=blue,
  pdftitle={Sparse Recovery With Integrality Constraints},
  pdfauthor={J.-H. Lange, M. E. Pfetsch, B. M. Seib, A. M. Tillmann}]{hyperref}

\newcommand{\R}{\mathds{R}}
\newcommand{\N}{\mathds{N}}
\newcommand{\Z}{\mathds{Z}}
\newcommand{\Q}{\mathds{Q}}

\newcommand{\vones}{\boldsymbol{\mathds{1}}}

\newcommand{\norm}[1]{\lVert {#1} \rVert}
\newcommand{\abs}[1]{\lvert {#1} \rvert}
\newcommand{\card}[1]{\lvert {#1} \rvert}
\newcommand{\suchthat}{\, : \,}
\newcommand{\define}{\coloneqq}
\newcommand{\ceil}[1]{\lceil{#1}\rceil}
\newcommand{\floor}[1]{\lfloor{#1}\rfloor}
\newcommand{\T}{^\top}
\newcommand{\cN}{\mathcal{N}}
\newcommand{\NP}{\textsf{NP}}
\newcommand{\coNP}{\textsf{coNP}}
\renewcommand{\ker}{\mathcal{N}}

\DeclareMathOperator{\range}{range}
\DeclareMathOperator{\rank}{rank}
\DeclareMathOperator{\supp}{supp}
\DeclareMathOperator{\sign}{sign}
\DeclareMathOperator{\spark}{spark}
\DeclareMathOperator{\conv}{conv}
\DeclareMathOperator{\nspop}{NSP}

\newcommand{\Prob}[2]{(\text{\normalfont P}_{#1}({#2}))}
\newcommand{\ProbLP}[2]{(\text{\normalfont P}^{\text{\normalfont LP}}_{#1}({#2}))}

\newcommand{\Pone}{\Prob{1}{\Z^n}}

\newcommand{\PoneN}{\Prob{1}{\Z_+^n}}

\newcommand{\PoneP}{\Prob{1}{[\0,\u]_\Z}}
\newcommand{\PzeroPM}{\Prob{0}{[\l,\u]_\Z}}
\newcommand{\PonePM}{\Prob{1}{[\l,\u]_\Z}}

\newcommand{\NSP}[1]{\nspop(#1)}
\newcommand{\NSPP}[1]{\nspop_+(#1)}

\newcommand{\A}{\bm{A}}

\newcommand{\B}{\bm{B}}

\newcommand{\I}{\bm{I}}

\newcommand{\U}{\bm{U}}
\newcommand{\V}{\bm{V}}
\renewcommand{\l}{\bm{\ell}}
\renewcommand{\a}{\bm{a}}
\renewcommand{\b}{\bm{b}}
\renewcommand{\u}{\bm{u}}
\renewcommand{\c}{\bm{c}}
\renewcommand{\v}{\bm{v}}
\newcommand{\w}{\bm{w}}
\newcommand{\x}{\bm{x}}
\newcommand{\y}{\bm{y}}
\newcommand{\z}{\bm{z}}
\renewcommand{\b}{\bm{b}}
\newcommand{\0}{\bm{0}}
\newcommand{\bdelta}{\bm{\delta}}

\newtheorem{theorem}{Theorem}[section]
\newtheorem{lemma}[theorem]{Lemma}
\newtheorem{corollary}[theorem]{Corollary}
\newtheorem{remark}[theorem]{Remark}
\newtheorem{example}[theorem]{Example}
\newtheorem{proposition}[theorem]{Proposition}
\newtheorem{definition}[theorem]{Definition}

\begin{document}

\title{Sparse Recovery With Integrality Constraints}
%

\author{Jan-Hendrik Lange}\ead{jlange@mpi-inf.mpg.de}\address{Max-Planck-Institute for Informatics, Campus E1 4, 66123 
  Saarbr\"ucken, Germany}
\author{Marc E. Pfetsch}\ead{pfetsch@mathematik.tu-darmstadt.de}
\author{Bianca M. Seib\fnref{fn1}}\ead{bianca.seib@gmail.com}\address{Research Group Optimization, TU Darmstadt, Dolivostr.\ 15, 64293 Darmstadt, 
  Germany}
\author{Andreas M. Tillmann\corref{cor1}}\ead{tillmann@or.rwth-aachen.de}\address{Visual Computing Institute \& Operations Research
  Group, RWTH Aachen University, Lehrstuhl f\"ur Informatik~8, Ahornstr.\
  55, 52074 Aachen, Germany}

\cortext[cor1]{Corresponding author}
\fntext[fn1]{Present address: Wiesbaden, Germany}

\markboth
{Sparse Recovery With Integrality Constraints}
{Sparse Recovery With Integrality Constraints}

\begin{abstract}
  \noindent
  We investigate conditions for the unique recoverability of sparse
  integer-valued signals from a small number of linear measurements. Both
  the objective of minimizing the number of nonzero components, the
  so-called $\ell_0$-norm, as well as its popular substitute, the
  $\ell_1$-norm, are covered. Furthermore, integrality constraints and
  possible bounds on the variables are investigated. Our results show that
  the additional prior knowledge of signal integrality allows for
  recovering more signals than what can be guaranteed by the established
  recovery conditions from (continuous) compressed sensing. Moreover, even
  though the considered problems are \NP-hard in general (even with an
  $\ell_1$-objective), we investigate testing the $\ell_0$-recovery
  conditions via some numerical experiments. It turns out that the
  corresponding problems are quite hard to solve in practice using
  black-box software. However, medium-sized instances of $\ell_0$- and
  $\ell_1$-minimization with binary variables can be solved exactly within
  reasonable time. 
\end{abstract}


\maketitle

\section{Introduction}\label{sec:intro}

\noindent
The recovery of sparse signals has received a tremendous interest in recent
years. The basic setting without noise is as follows: under the prior
knowledge that a measurement vector $\b
\in \R^m \setminus \{0\}$ is generated by a sparse signal $\x \in
\R^n$ via $\A \x = \b$, where $\A \in \R^{m \times n}$ with
$\rank(\A)=m < n$ is the sensing matrix, the question is whether $\x$
can be uniquely recovered, given $\A$ and $\b$. Thus, one approach is
to find the sparsest $\x$ that explains the measurements, i.e., one
minimizes $\norm{\x}_0 \define \card{\{i \in \{1, \dots, n\} \suchthat
  x_i \neq 0\}}$ under the constraint $\A\x=\b$. However, this problem
is \NP-hard, see Garey and Johnson~\cite{GarJ79}. The crucial idea in
this context (see, e.g., Chen et al.~\cite{CheDS98}) is to replace
$\norm{\x}_0$ by the $\ell_1$-norm $\norm{\x}_1 \define \abs{x_1} +
\dots + \abs{x_n}$, which results in a convex problem that can even be
cast as a linear program (LP) and is therefore tractable. The
literature contains an abundance of conditions under which minimizers
of $\norm{\x}_1$ subject to $\A \x = \b$ are unique and equal to the
sparsest solution; at this place, we refer to the book by Foucart and
Rauhut~\cite{FouR13} for more information and an overview of selected
specialized algorithms to solve the $\ell_1$-minimization problem.

The key point for the mentioned series of striking results is the
prior knowledge that~$\b$ can be sparsely represented or
approximated. A natural question is whether further knowledge about
the structure of the representations~$\x$ can lead to stronger results
about the recoverability. In general terms, the two problems from
above can be written as
\begin{align*}
  \min\; \{\norm{\x}_0 \suchthat \A \x = \b,\; \x \in X\},\label{prob:P0}\tag{$\text{P}_0(X)$}\\
  \min\; \{\norm{\x}_1 \suchthat \A \x = \b,\; \x \in X\},\label{prob:P1}\tag{$\text{P}_1(X)$}
\end{align*}
where $X \subseteq \R^n$ is a constraint set representing further
restrictions on the representations.

The ``classical'' results in the literature refer to the case $X = \R^n$.
One main example in which $X \neq \R^n$ is the case in which~$\x$ has to be
nonnegative, i.e., $X = \R_+^n$, see, for instance, Donoho and Tanner~\cite{DonT05},
Bruckstein et al.~\cite{BruEZ08}, and Khajehnejad et al.~\cite{KhaDXH11}.

In this paper, we investigate the case in which $\x$ is required to be
\emph{integral}, i.e., $X \subseteq \Z^n$. Thus, we investigate the
interplay (and tradeoff) of the prior information of sparsity and
integrality of the solution. This setting is motivated by various
applications in which signals are composed from finite symbol alphabets,
such as machine-to-machine communication (see Knopp et
al.~\cite{KnoMBPDPD14}), spectrum sensing for cognitive radio (cf.\ Axell et
al.~\cite{AxeLLP12}), detection, localization or interference cancellation
in multiple-input/output (MIMO) systems (see, e.g., Zhu and
Giannakis~\cite{ZhuG11}, Knopp et al.~\cite{KnoMBWPD13} and Rossi et
al.~\cite{RosHE14}), or discrete tomography tasks as described by Batenburg
and Sijbers~\cite{BatS11} or Kuske et al.~\cite{KusSP17}, to name but a
few.

One particular application arises when constellation signals are used in
massive MIMO; we briefly describe the real-valued case, for
simplicity. Here, the components of the signal are chosen from a small set
of constellation signals $\{C_1, \dots, C_M\}$. One class of examples is
given by the $M$ phase-shift keying ($M$-PSK); several different types of
such configurations with different values of $M$ exist. Then, one can
multiply the columns of the sensing matrix with all constellation signals,
and binary variables $\x$ can be used to select the corresponding signal
for each component. If one searches for the sparsest signal vector (e.g.,
in the context of low-activity multi-user detection considered
in~\cite{KnoMBPDPD14}), one arrives at an instance of
problem~\eqref{prob:P0} in the noise-free case. Hegde et
al.~\cite{HedYSP16} discuss an optimization approach to relax such signals,
Hegde et al.~\cite{HedPP17} propose an exact method, and, e.g.,
Zhu and Giannakis~\cite{ZhuG11} treats binary PSK.

\begin{example}\label{ex:basic}
  To illustrate some of the issues investigated in this paper, consider
  $\A = (2, 3, 6) \in \R^{1 \times 3}$ and $\b = (11)$. Then
  $\Prob{0}{\Z_+^3}$ has the two optimal solutions $(4, 1, 0)\T$ and
  $(1, 3, 0)\T$. Furthermore, $\Prob{1}{\Z_+^3}$ has optimal solution
  $(1,1,1)\T$. Finally, $\Prob{0}{\R_+^3}$ has three optimal solutions,
  each with one nonzero entry, while $\Prob{1}{\R_+^3}$ has the unique
  optimal solution $(0,0,\tfrac{11}{6})\T$. This shows that requiring
  integrality affects the optimal solutions of~\eqref{prob:P0}
  and~\eqref{prob:P1}. Moreover, these problems may yield different
  solutions.
\end{example}

Despite the apparent practical interest, there are only a few articles
in the literature that deal with integral signal recovery, both
theoretically and algorithmically. For instance, Sparrer and
Fischer~\cite{SpaF16} present a heuristic approach based on orthogonal
matching pursuit. A further heuristic was proposed by Flinth and
Kutyniok~\cite{FliK16} based on a combination of projection and
orthogonal matching pursuit ideas. The binary case---particularly
prominent in the context of digital/wireless communication systems,
where transmitted signals can often be represented as simple bit
sequences---has been treated, for instance, by Nakarmi and
Rahnavard~\cite{NakR12} and Wu et al.~\cite{WuFLZ09}. Mangasarian and
Recht~\cite{ManR11} gave conditions for uniqueness of vectors in
$X=\{-1,1\}^n$ as solutions of $\A\x=\b$, $-\vones\leq\x\leq\vones$,
while Stojnic~\cite{Sto10} determined empirical and probabilistic
theoretical recovery thresholds for binary signals
via~$\Prob{1}{[0,1]^n}$. Swoboda et al.~\cite{SwoKS16} presented a
Lagrangian relaxation based heuristic for solving problems with
integer variables.

The above-mentioned works mostly exhibit a clear focus on the algorithmic
side, proposing relaxations or heuristic methods and empirical results on
their success. Rigorous theoretical conditions for the recovery of sparse
integral signals by $\ell_1$-minimization with relaxed integrality
requirements were presented in Keiper et al.~\cite{KeiKLP17}. They
investigate binary and ternary signals by means of $\Prob{1}{[0,1]^n}$ and
$\Prob{1}{[-1,1]^n}$. One of their main contributions is the investigation
of phase transitions of unique recovery, which showed that recovery
exploiting the bounds $[0,1]$ or $[-1,1]$ takes place earlier. Furthermore,
Flinth and Keiper~\cite{FliK18} provided a more detailed investigation of
binary signal recovery by providing probabilistic conditions for unique
recovery.

Our work complements those results and generalizes some of them: We
consider more general integral sets and their continuous relaxations along
with both $\ell_0$- and $\ell_1$-objectives. For instance, one of our
results shows that explicitly treating integrality constraints can allow
for the recovery of essentially arbitrarily many more integral signals than
could be recovered by the associated relaxed (integrality-oblivious)
problem:
\begin{example}\label{ex:boundedNSPsparsityDiff1}
  Let $-\l=\u=2\cdot\vones\in\Z^n$ with $n\geq 10$, let
  $\v=(2,-2,(\v^\prime)^\top)^\top$ and
  $\w=(\floor{\tfrac{n}{2}}, \floor{\tfrac{n}{2}}, (\w^\prime)^\top)^\top$
  with $\v^\prime\in\{-2,2\}^{n-2}$ and $\w^\prime\in\{-1,1\}^{n-2}$
  arbitrary. Let $\A$ be such that its nullspace is
  $\text{\emph{span}}\{\v,\w\}$. Then Proposition~\ref{thm:boundedNSPsuff}
  yields recoverability of all
  $\x\in X\define\{\x\in\Z^n \suchthat \l\leq\x\leq\u\}$ with at most
  $(\floor{\tfrac{n}{2}} - 1)$ nonzeros by means of $\Prob{1}{X}$. However,
  ignoring integrality, which amounts to solving $\Prob{1}{\conv(X)}$,
  one cannot be sure to recover all sparse signals from $X$ with as few as
  $2$ nonzero components; see Example~\ref{ex:boundedNSPsparsityDiff2} in
  Section~\ref{subsec:l1nsps} for the details.
\end{example}

The computational price one has to pay for such strong results is that the
$\ell_0$- and $\ell_1$-problems are \NP-hard if integrality of the signals
is enforced, see Section~\ref{sec:basics}. Therefore, the main motivation
for this paper is to provide a comprehensive theoretical characterization
of cases in which it is worth investing additional computational resources
to take integrality into account. Our results thus also provide a
motivation for developing specialized heuristic or exact solution
techniques. (Indeed, some of the computational results we will present
demonstrate that the problems under consideration may be very hard to solve
with general-purpose black-box algorithms, emphasizing the need to develop
problem-specific methods in the future.)

It is important to note that we only treat the noise-free case, i.e., we
consider equations $\A \x = \b$ instead of an error bound like
$\norm{\A \x - \b}_2 \leq \delta$. Thus, our results mark a first step
towards an understanding of the underlying structure and should be extended
to the noise-aware case to be more relevant for real-world applications in
the future.

The remainder of the paper is organized as follows. In
Section~\ref{sec:basics}, after formally specifying the problems treated in
the paper, we first show some basic \NP-hardness results and that choosing
rational $\A$ may have a crucial impact on the recoverability properties,
and then provide an overview of our main contributions. In
Section~\ref{sec:l0case}, we derive recoverability characterizations for
the $\ell_0$-problems. In Section~\ref{sec:l1case}, we turn to the
$\ell_1$-case and derive characterizations for uniform
(Section~\ref{subsec:l1nsps}) and individual
(Section~\ref{subsec:P1indivRec}) recoverability. An additional discussion
of cases in which an integral solution can be guaranteed when solving the
continuous relaxation (without aiming at solution uniqueness or even
sparsity) is provided in Appendix~\mbox{\ref{subsec:unimodTDIetc}}. In
Section~\ref{sec:experiments}, we report on some computational experiments,
and close with final remarks in Section~\ref{sec:coda}.

\begin{remark}\label{rem:complexSignals}
  Many of the main results in compressed sensing also hold with
  respect to complex data and signals, see, e.g., the survey of real
  and complex nullspace conditions characterizing
  $\ell_0$-$\ell_1$-equivalence in~\cite{FouR13}. Nevertheless, for
  the sake of simplicity, we only consider the real-valued case in
  this paper. For instance, extensions to complex signals with (say)
  integral real and imaginary parts are not treated here.
\end{remark}

We use the following notation: We use $\N = \{1, 2, \dots\}$ and
define $[n] \define \{1, \dots, n\}$ for $n \in \N$. Furthermore, for
$s \in [n]$, the vector $\x \in \R^n$ is \emph{$s$-sparse}, if
$\norm{\x}_0 \leq s$. The \emph{support} of $\x$ is defined as
$\supp(\x) \define \{i \in [n] \suchthat x_i \neq 0\}$. Moreover, for
$S \subseteq [n]$, $\x_S \in \R^n$ denotes the vector which equals
$\x$ for all components indexed by~$S$ and is $0$ otherwise. The
complement of a set~$S \subseteq [n]$ is denoted by $S^c \define [n]
\setminus S$. The nullspace (kernel) of a matrix $\A$ is defined as
$\ker(\A)\define\{\x\suchthat\A\x=\0\}$. By $\vones$, we denote the
all-ones vector of appropriate dimension. 

\section{Overview, Contributions and Basic Results}\label{sec:basics}

\noindent
In this paper, we investigate the following five basic integrality requirements
for \eqref{prob:P0} and \eqref{prob:P1}: 
\begin{equation}\label{eq:BasicX}
  X = \Z^n,\quad X = \Z^n_+,\quad X = [-\u, \u]_{\Z},\quad X = [\0, \u]_{\Z},\quad X = [\l, \u]_{\Z},
\end{equation}
where $\l \leq \0 \leq \u \in \R^n$ and
$[\l, \u]_\Z \define \{\x \in \Z^n \suchthat \l \leq \x \leq \u\}$. Note
that we have to make sure that $\0 \in X$ in order to allow sparse
solutions; moreover, throughout the paper, we assume without loss of
generality that $\l<\u$ (otherwise, $\ell_i = u_i=0$, so $x_i=0$ can be
eliminated from the problem a priori). When considering $[\l, \u]_\Z$, we
can round the components of $\l$ and $\u$ up and down, respectively; thus,
in this case, we may assume that $\l$, $\u \in \Z^n$. However, in
particular cases, we also deal with boxes
$[\l, \u]_\R \define \{\x \in \R^n \suchthat \l \leq \x \leq \u\}$ for
which $\l$ and $\u$ can be real-valued. Clearly, $[\l, \u]_\Z$ is the most
general (integral) case; the others can be written in this form (if $\l$
and $\u$ are allowed to take $\mp \infty$ values, respectively).

The first observation is that all of the considered problems are \NP-hard.

\begin{proposition}\label{prop:complexity}
  The problems \eqref{prob:P0} and \eqref{prob:P1} are \NP-hard in the
  strong sense for each of the sets~$X$ in~\eqref{eq:BasicX}, even if
  $\A$ is binary and $\b = \vones$.
\end{proposition}
\begin{proof}
  Garey and Johnson~\cite{GarJ79} proved that $\Prob{0}{\R^n}$ is
  strongly \NP-hard using a reduction from ``exact cover by 3-sets''
  (this proof is reproduced in~\cite{FouR13}). The proof shows that,
  given an instance of this problem, one can construct a binary matrix
  $\A$ such that solutions~$\x$ of $\A \x = \vones$ minimizing
  $\norm{\x}_0$ are necessarily $0/1$, i.e., $\x \in \{0,1\}^n$. These
  solutions are feasible for any of the considered problems and
  furthermore satisfy $\norm{\x}_0 = \norm{\x}_1$.
\end{proof}
\medskip

This result carries an unfortunate negative message: Unlike in the
real-valued setting, changing $\norm{\x}_0$ to $\norm{\x}_1$ does not
change the complexity status of the problem, and all considered problems
are hard to solve. On the other hand, modern integer optimization
technology allows to solve small to medium-sized instances of these
problems. Moreover, empirically, the $\ell_1$-case is often slightly
easier.

In any case, the question to what extent integrality requirements allow to
increase the number of cases in which a signal can be uniquely recovered is
fundamental. Such solutions might then be found efficiently in
practice, e.g., by heuristics such as that by Flinth and
Kutyniok~\cite{FliK16}.  \medskip

When considering integrality requirements, it is of fundamental importance
whether the matrix $\A$ is rational:

\begin{proposition}\label{thm:IrrationalRecovery}
  For any $n \in \N$, there exists a single-row matrix $\A \in \R^{1 \times n}$ such
  that for every $\b \in \range_{\Z}(\A) \define \{\A \z \suchthat \z \in
  \Z^n\}$, there exists a unique $\x \in \Z^n$ such that $\A \x = \b$.
\end{proposition}
\begin{proof}
  Since $\R$ is an infinite-dimensional vector space over $\Q$,
  choosing $n$ real numbers that are linearly independent over $\Q$ as
  the components of $\A$ suffices. For instance, taking the $n$th
  roots of pairwise different prime numbers $\geq 2$ will do, see,
  e.g., Besicovitch~\cite{Bes40}. Thus, $\A \x = \b$ has a unique
  integral solution for every $\b \in \range_{\Z}(\A)$.
\end{proof}
\medskip

As a consequence, for such a matrix $\A$, the recovery problem with
integral~$\x$ is always uniquely solvable and thus, ideal recovery is
possible. However, in general, such matrices cannot be stored in a
computer. Moreover, for general real-valued $\A$ (with irrational entries),
the complexity of solving $\A \x = \b$ with $\x \in \Z^n$ is unclear; in
particular, one needs to use a ``non-standard'' model of computation, cf.,
e.g., Blum et al.~\cite{BluCSS97}.

In the following, we will often consider \emph{rational} $\A$. Note
that in this case, finding \emph{some} integral solution~$\x$ of $\A
\x = \b$ can be done in polynomial time using the Hermite normal form,
see, e.g., Schrijver~\cite{Sch86}. However, as
Proposition~\ref{prop:complexity} shows, minimizing $\norm{\x}_0$ or
$\norm{\x}_1$ is still \NP-hard.

\subsection{Main Contributions}\label{subsec:contributions}

\noindent
As mentioned in the introduction, the main purpose of this work is to
provide a comprehensive treatment of conditions that guarantee successful
recovery of sparse integral vectors from underdetermined linear equation
systems. By considering every possible case, comparing with the
corresponding continuous settings and pointing out interesting subtleties
(such as differences that may occur depending on whether $\A$ is real or
rational), we hope to lay a thorough theoretical foundation for integral
sparse signal recovery, and coincidentally close some gaps in the existing
literature for the continuous setting as well. For completeness, we also
include relevant known results. Moreover, for some select example problems,
we take a first step towards their practical solution and
evaluating the derived recovery conditions, by formulating integer
programming models that can be handled by state-of-the-art general-purpose
mixed-integer programming software. One important take-away message from
the computational experiments is that to efficiently solve medium- to
large-scale instances in practice, one will likely need to develop
problem-specific specialized solution methods (which is out of scope of the
present work). Since the gains with respect to integral signal
recoverability provided by our theory can be quite significant, such
algorithm design endeavors as well as generally spending more computational
resources on the reconstruction process are well-justified.

Sections~\ref{sec:l0case} and~\ref{sec:l1case} contain the detailed
discussions of recoverability by means of solving \eqref{prob:P0} and
\eqref{prob:P1} for each of the sets $X$ specified in~\eqref{eq:BasicX} and
the associated continuous relaxations, respectively. To help navigate the
somewhat long series of results, Table~\ref{tab:mainresults} gives an
overview of the main results. (We omit the cases $X=[\0,\u]_\Z$ and
$X=[-\u,-\u]_\Z$ (and their relaxations) from the table, since they are
special cases of the most general one, $X=[\l,\u]_\Z$.) The table lists the
different problems, the conditions characterizing \emph{uniform} recovery
(i.e., of all $s$-sparse signals from the respective sets)---precise
definitions can be found in Sections~\ref{sec:l0case} and~\ref{sec:l1case},
respectively---as well as references to the corresponding results in this
work or, where applicable, from the existing literature.

\begin{table}[t]
  \caption{Overview of problems and conditions for uniform
    continuous or discrete $s$-sparse signal recovery. $S_{\pm}(\z)$ denotes supports of positive/negative entries of
    $\z$, respectively.}
  \label{tab:mainresults}
  \vspace*{-0.75em}
  \begin{center}
    \footnotesize
    \begin{tabular}{@{}l@{\quad\,\,}c@{\quad\,\,}l@{}}\toprule
      problem & uniform recovery condition (order~$s$) & reference\\\midrule
      P$_0(\R^n)$       & $\cN(\A)\cap\{\z\in\R^n\suchthat\norm{\z}_0\leq 2s\}=\{\0\}$ & \cite{CohDV09,FouR13} (cf. Thm.~\ref{thm:R-s-good})\\
      P$_0(\R_+^n)$     & $\cN(\A)\cap\{\z\in\R^n\suchthat\abs{S_{-}(\z)}\leq s,~\abs{S_{+}(\z)}\leq s\}=\{\0\}$ & this work (Cor.~\ref{cor:P0Real}) \\      
      P$_0([\l,\u]_\R)$ & $\cN(\A)\cap C_\R(\l,\u)=\{\0\}$ & this work (Cor.~\ref{cor:P0Real})\\[0.5em]
      P$_1(\R^n)$       & NSP($\R^n$) & (e.g.)~\cite{FouR13} (cf. Thm.~\ref{thm:nspsReal})\\
      P$_1(\R_+^n)$     & NSP$_+$($\R^n$) & \cite{KhaDXH11} (cf. Thm.~\ref{thm:nspsReal})\\
      P$_1([\l,\u]_\R)$ & NSP($\R^n$) & this work (Cor.~\ref{cor:luNSPRequivNSPR})\\[0.5em]
      P$_0(\Z^n)$       & $\cN(\A)\cap\{\z\in\Z^n\suchthat\norm{\z}_0\leq 2s\}=\{\0\}$ & this work (Thm.~\ref{thm:Z-s-good})\\
      P$_0(\Z_+^n)$     & $\cN(\A)\cap\{\z\in\Z^n\suchthat\abs{S_{-}(\z)}\leq s,~\abs{S_{+}(\z)}\leq s\}=\{\0\}$ & this work (Cor.~\ref{cor:recovery})\\
      P$_0([\l,\u]_\Z)$ & $\cN(\A)\cap C_\Z(\l,\u)=\{\0\}$ & this work (Thm.~\ref{thm:P0Master})\\[0.5em]
      P$_1(\Z^n)$       & NSP($\Z^n$) & this work (Thm.~\ref{thm:nspsZuZp})\\
      P$_1(\Z_+^n)$     & NSP$_+$($\Z^n$) & this work (Thm.~\ref{thm:nspsZuZp})\\
      P$_1([\l,\u]_\Z)$ & $(\A,-\A)$ satisfies $\displaystyle{\text{NSP}_{+}\left(\left[\left(\begin{subarray}{c}-\u\\\l\end{subarray}\right),\left(\begin{subarray}{c}\u\\-\l\end{subarray}\right)\right]_{\Z}\right)}$ & this work (Thm.~\ref{thm:PonePMnsp})\\\bottomrule
    \end{tabular}
  \end{center}
\end{table}

\section{The $\ell_0$-case}\label{sec:l0case}

\noindent
In this section, we provide conditions on the uniform recoverability
via~\eqref{prob:P0}. For this, we define the set
\[
S(s, X; \b) \define \{\x \suchthat \A \x = \b,\; \norm{\x}_0 \leq s,\; \x \in X\}.
\]
The key point is uniqueness of sparse solutions, i.e., whether
$\card{S(s, X; \A \hat{\x})} = 1$ for $s$-sparse $\hat{\x} \in
X$. Inspired by the terminology of Juditsky and
Nemirovski~\cite{JusN11}, we define the following.  \smallskip

\begin{definition}\label{def:0-good}
  Let $\A \in \R^{m \times n}$, $s \in [n]$, and $X \subseteq
  \R^n$. The matrix $\A$ is \emph{$(s,X,0)$-good}, if for \emph{every} $s$-sparse
  vector $\hat{\x} \in X$, \mbox{$\card{S(s, X; \A \hat{\x})} = 1$} holds.
\end{definition}

We first state some obvious results for $(s,X,0)$-good matrices.

\begin{lemma}\label{lem:SgoodProp}
  Let $\A \in \R^{m \times n}$, $s \in [n]$, and $X \subseteq \R^n$.
  \begin{enumerate} 
  \item If $\A$ is $(s,X,0)$-good, it is $(s,X',0)$-good for every $X'
    \subseteq X$. Therefore,
    \[
    \text{$(s,\R^n,0)$-good}
    \quad\Rightarrow\quad
    \text{$(s,\Z^n,0)$-good}
    \quad\Rightarrow\quad
    \text{$(s,[\l,\u]_\Z,0)$-good}.
    \]
    Moreover, if $\A$ is $(s,[\l,\u]_\Z,0)$-good, it is
    $(s,[\l',\u']_\Z,0)$-good for every $\l \leq \l' \leq \0 \leq \u' \leq \u$ ($\l,\u\in\R^n$).
  \item If $\A$ is $(s,X,0)$-good, it is $(s',X,0)$-good for every $s' \leq s$,
    $s' \in \N$.
  \end{enumerate}
\end{lemma}

Furthermore, we recall the following well-known result from the literature.

\begin{theorem}[\cite{CohDV09}, \cite{FouR13}]\label{thm:R-s-good}
  A matrix $\A \in \R^{m \times n}$ is $(s,\R^n,0)$-good for $s \in [n]$ if
  and only if $\ker(\A) \cap \{\z \in \R^n \suchthat \norm{\z}_0 \leq 2s\}
  = \{\0\}$.
\end{theorem}

The statement of this theorem can be rephrased by using $\spark(\A)
\define \min\,\{\norm{\x}_0 \suchthat \A \x = \0,\; \x \neq \0\}$, which
refers to the smallest number of linearly dependent columns
of~$\A$. Then, $\A$ is $(s,\R^n,0)$-good if and only if $\spark(\A) >
2s$. Since the decision problem ``is $\spark(\A)\leq k$\,?'' is \NP-complete
(cf.\ \cite{TilP14}) and a $\z\in\Q^n$ with
$1\leq\norm{\z}_0\leq 2s$ serves as a certificate for
$\A\in\Q^{m\times n}$ \emph{not} being $(s,\R^n,0)$-good, this shows
that checking the condition in Theorem~\ref{thm:R-s-good} is
\coNP-complete.

By a completely analogous proof, Theorem~\ref{thm:R-s-good} carries over to
the integral case by requiring $\z \in \Z^n$. Moreover,
if $\A$ is rational, we can always scale vectors in the nullspace $\ker(\A)$
to be integral. This yields: 

\begin{theorem}\label{thm:Z-s-good}
  A matrix $\A \in \R^{m \times n}$ is $(s,\Z^n,0)$-good if and only if
  $\ker(\A) \cap \{\z \in \Z^n \suchthat \norm{\z}_0 \leq 2s\} = \{\0\}$. Thus,
  if $\A \in \Q^{m \times n}$, then $\A$ is $(s,\Z^n,0)$-good if and
  only if it is $(s,\R^n,0)$-good.
\end{theorem}

Again using \NP-completeness for the spark, checking the condition in
Theorem~\ref{thm:Z-s-good} is also \coNP-complete.  Moreover,
Theorem~\ref{thm:Z-s-good} has the following interesting consequence,
compare with Proposition~\ref{thm:IrrationalRecovery}.

\begin{corollary}
  Let $s \in [n]$. The minimal number of rows $m$ for which a
  \emph{rational} matrix $\A \in \Q^{m \times n}$ can be $(s,\Z^n,0)$-good is $2s$.
\end{corollary}
\begin{proof}
  If $\A$ is rational, the condition in Theorem~\ref{thm:Z-s-good} is
  equivalent to that of Theorem~\ref{thm:R-s-good}. Moreover, for continuous settings, 
  \cite[Theorems~2.13 and~2.14]{FouR13} (see also Cohen et al.~\cite{CohDV09})
  show that $m \geq 2\, s$ is necessary in general and equality can be
  achieved using a Vandermonde matrix.
\end{proof}
\medskip

On the other hand, when additionally considering bounds on the
variables, we get a similar behavior as in
Proposition~\ref{thm:IrrationalRecovery} even for rational matrices:

\begin{proposition}\label{thm:BoundedRecovery}
  Let $X = [\0,\u]_\Z$ for $\u \in \Z_{>0}^n$. Then for any $n \in \N$
  there exists a rational matrix $\A \in \Q^{1 \times n}$ such that for
  every $\b \in \range_{X}(\A) \define \{\A \z \suchthat \z \in X\}$ there
  exists a unique $\x \in X$ such that $\A \x = \b$.
\end{proposition}
\begin{proof}
  Define $\delta \define \max\,\{u_1, \dots, u_n\} + 1$ and let $A =
  (a_{1k})$ be defined by $a_{1k} \define \delta^k$ for $k = 0, \dots,
  n-1$. Then, $\b = \A \x$ for $\x \in X$ amounts to a $\delta$-ary
  expansion of $\b$, which is unique.
\end{proof}
\medskip

Note that this result is of theoretical interest only, since the large
coefficients from the proof of Proposition~\ref{thm:BoundedRecovery} will
produce numerical problems for larger~$n$.

\subsection{Recovery Conditions for the $\ell_0$-Case}

\noindent
To treat the case of $\PzeroPM$, we need the following notation.  For $\a
\leq \b \in \R^n$, we consider closed boxes $[\a,\b] \define \{\v \suchthat
\a \leq \v \leq \b\} = [a_1, b_1] \times \dots \times [a_n,
b_n]$; similarly, half-open boxes are defined in the obvious way. For $\z
\in \R^n$ and one of these boxes $B = B_1 \times \dots \times B_n \subseteq \R^n$, we
define $\supp(\z; B) \define \{i \in [n] \suchthat z_i \in B_i\}$.

\begin{theorem}\label{thm:P0Master}
  Let $X = [\l,\u]_\Z$, $\delta^{\min}_i \define \min \{-\ell_i, u_i\}$
  and $\delta^{\max}_i \define \max\{-\ell_i,u_i\}$ for all
  $i\in[n]$. Furthermore, define the following (possibly empty) sets
  depending on a vector $\z \in \R^n$
  \begin{align*}
    S_1^+ & \define \supp(\z; (\0, \bdelta^{\min}]),    & S_1^- & \define \supp(\z; [-\bdelta^{\min},\0)), \\
    S_2^+ & \define \supp(\z; (\bdelta^{\min}, \u]),    & S_2^- & \define \supp(\z; [\l, -\bdelta^{\min})), \\
    S_3^+ & \define \supp(\z; (\u, \bdelta^{\max}]),    & S_3^- & \define \supp(\z; [-\bdelta^{\max}, \l)), \\
    S_4^+ & \define \supp(\z; (\bdelta^{\max}, \u-\l]), & S_4^- & \define \supp(\z; [\l-\u,\, -\bdelta^{\max}))
 \end{align*}
 and let
 \begin{align*}
   C_\Z(\l,\u) \define \big\{\z \in [\l-\u,\u-\l]_\Z \suchthat
    \card{S_4^+} + \card{S_4^-} + \card{S_3^+} + \card{S_3^-} \leq s&,\\ 
    \card{S_4^+} + \card{S_4^-} + \card{S_2^+} + \card{S_2^-} \leq s&,\\
    2\, \big(\card{S_4^+} + \card{S_4^-}\big) + \card{S_3^+} + \card{S_3^-} + \card{S_2^+} + \card{S_2^-} + \card{S_1^+} + \card{S_1^-} \leq 2s&\big\}.
 \end{align*}
 Then, an $\A \in \R^{m \times n}$ is $(s,[\l,\u]_\Z,0)$-good if and
 only if $\ker(\A) \cap C_\Z(\l,\u) = \{\0\}$.
\end{theorem}

\begin{proof}
  Without loss of generality, we assume that $\l,\u\in\Z^n$. We first observe that
  \[
  \l - \u \leq -\bdelta^{\max} \leq \l \leq -\bdelta^{min} \leq \0 \leq
  \bdelta^{\min} \leq \u \leq \bdelta^{\max} \leq \u - \l,
  \]
  see also Figure~\ref{fig:P0Master}. This shows that the boxes on which
  the sets $S_1^+$ to $S_4^-$ are based are well-defined, although
  they may be empty.
  \begin{figure}[tb]
    \begin{center}
      \begin{tikzpicture}[scale=0.69]
        \footnotesize
        \tikzstyle{node} += [draw=black, circle, fill=black, inner sep=2pt];
        \tikzstyle{arc} += [draw=black,thick];
        
        \node (lum)   at (0,0) [node,label=below:$\ell_i - u_i$] {};
        \node (dmaxm) at (2,0) [node,label=below:$-\delta^{\max}_i$] {};
        \node (lm)    at (4,0) [node,label=below:$\ell_i$] {};
        \node (dminm) at (6,0) [node,label=below:$-\delta^{\min}_i$] {};
        
        \node (zero)  at (8,0) [node,label=below:$0$] {};
        
        \node (dminp) at (10,0) [node,label=below:$\delta^{\min}_i$] {};
        \node (up)    at (12,0) [node,label=below:$u_i$] {};
        \node (dmaxp) at (14,0) [node,label=below:$\delta^{\max}_i$] {};
        \node (ulp)   at (16,0) [node,label=below:$u_i - \ell_i$] {};
        
        \draw [arc] (lum) -- node[midway,above] {$S_4^-$} (dmaxm);
        \draw [arc] (dmaxm) -- node[midway,above] {$S_3^-$} (lm);
        \draw [arc] (lm) -- node[midway,above] {$S_2^-$} (dminm);
        \draw [arc] (dminm) -- node[midway,above] {$S_1^-$} (zero);
        
        \draw [arc] (zero) -- node[midway,above] {$S_1^+$} (dminp);
        \draw [arc] (dminp) -- node[midway,above] {$S_2^+$} (up);
        \draw [arc] (up) -- node[midway,above] {$S_3^+$} (dmaxp);
        \draw [arc] (dmaxp) -- node[midway,above] {$S_4^+$} (ulp);
        
        \draw [arc,<-] (-0.5,0) -- (lum);
        \draw [arc,->] (ulp) -- (16.5,0);
      \end{tikzpicture}
    \end{center}
    \caption{Illustration of the intervals in Theorem~\ref{thm:P0Master}.}
    \label{fig:P0Master}
  \end{figure}
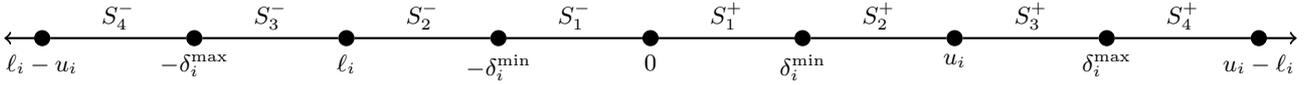

  Let $\A$ be $(s,[\l,\u]_\Z,0)$-good and let $\z \in \ker(\A) \cap
  C_\Z(\l,\u)$. Define $k \define \card{S_4^+} + \card{S_4^-} +
  \card{S_3^+} + \card{S_3^-} \leq s$ and $r \define \card{S_1^+} +
  \card{S_1^-}$.  Let $\tilde{S}$ be composed of $\min\,\{r, s - k\}$
  arbitrary indices of $S_1^+ \cup S_1^-$ and $\tilde{S}^c \define
  (S_1^+ \cup S_1^-) \setminus \tilde{S}$ be its complement
  (with respect to $S_1^+\cup S_1^-$). Now, we define
  \[
  x_i \define
  \begin{cases}
    \ell_i & \text{if } i \in S_4^-,\\
    0      & \text{if } i \in S_3^-,\\
    z_i    & \text{if } i \in S_2^-,\\
    0      & \text{if } i \in \tilde{S},\\
    z_i    & \text{if } i \in \tilde{S}^c,\\
    z_i    & \text{if } i \in S_2^+,\\
    0      & \text{if } i \in S_3^+,\\
    u_i    & \text{if } i \in S_4^+,\\
    0      & \text{otherwise},
  \end{cases}
  \quad\text{ and }\quad
  y_i \define
  \begin{cases}
    \ell_i - z_i & \text{if } i \in S_4^-,\\
    -z_i         & \text{if } i \in S_3^-,\\
    0            & \text{if } i \in S_2^-,\\
    -z_i         & \text{if } i \in \tilde{S},\\
    0            & \text{if } i \in \tilde{S}^c,\\
    0            & \text{if } i \in S_2^+,\\
    -z_i         & \text{if } i \in S_3^+,\\
    u_i - z_i    & \text{if } i \in S_4^+,\\
    0           & \text{otherwise}.
  \end{cases}
  \]
  These two vectors satisfy $\z = \x - \y$. Considering each case, one
  can see that $\x$, $\y \in X$.  Moreover,
  \[
  \card{\supp(\y)} = \card{S_4^+} + \card{S_4^-} + \card{S_3^+} +
  \card{S_3^-} + \card{\tilde{S}} = k + \min\,\{r, s - k\} \leq k + s - k = s.
  \]
  Furthermore, assume first that $r < s - k$, i.e., $\tilde{S} =
  S_1^+ \cup S_1^-$ and $\tilde{S}^c = \varnothing$. Then, by assumption,
  \[
  \card{\supp(\x)} = \card{S_4^+} + \card{S_4^-} + \card{S_2^+} +
  \card{S_2^-} \leq s.
  \]
  On the other hand, if $r \geq s -k$, then $\card{\tilde{S}} = s -k$ and
  $\card{\tilde{S}^c} = r - s + k$, which yields
  \begin{align*}
    & \card{\supp(\x)} = \card{S_4^+} + \card{S_4^-} + \card{S_2^+} + \card{S_2^-} + \card{\tilde{S}^c}\\
    =\,& \card{S_4^+} + \card{S_4^-} + \card{S_2^+} + \card{S_2^-} + (\card{S_1^+} + \card{S_1^-} - s + k)\\
    =\,& \card{S_4^+} + \card{S_4^-} + \card{S_2^+} + \card{S_2^-} + \card{S_1^+} + \card{S_1^-} - s + \card{S_4^+} +
    \card{S_4^-} + \card{S_3^+} + \card{S_3^-}\\
    =\,& 2\, (\card{S_4^+} + \card{S_4^-}) + \card{S_3^+} + \card{S_3^-} +
    \card{S_2^+} + \card{S_2^-} + \card{S_1^+} + \card{S_1^-} - s \leq s.
  \end{align*}
  Since $\z = \x - \y$, it follows that $\A \x = \A \y$ and consequently, by
  $(s,[\l,\u]_\Z,0)$-goodness of $\A$, that $\x = \y$, i.e., $\z = \0$.
  \smallskip

  Conversely, assume that $\ker(\A) \cap C_\Z(\l,\u) = \{\0\}$. Consider $\x,
  \tilde{\x} \in X$ with $\A \x = \A \tilde{\x}$, $\norm{\x}_0 \leq s$, and
  $\norm{\tilde{\x}}_0 \leq s$. By construction, $\z \define \x - \tilde{\x}
  \in \ker(\A) \cap [\l - \u, \u - \l]_\Z$.

  Now observe that if $i \in S_4^+$ then $x_i > 0$ and $\tilde{x}_i < 0$
  and similarly, if $i \in S_4^-$ then $x_i < 0$ and $\tilde{x}_i > 0$. This
  implies that $S_4^+ \subseteq \supp(\x) \cap \supp(\tilde{\x})$ and
  $S_4^- \subseteq \supp(\x) \cap \supp(\tilde{\x})$. For
  $S_4^c \define S_3^+ \cup S_3^- \cup S_2^+ \cup S_2^- \cup S_1^+
  \cup S_1^-$, we thus obtain
  \begin{align*}
    2 s  & \geq \card{\supp(\x)} + \card{\supp(\tilde{\x})}
    = \card{\supp(\x) \cup \supp(\tilde{\x})} + \card{\supp(\x) \cap \supp(\tilde{\x})}\\
    & \geq \card{\big(\supp(\x) \cup \supp(\tilde{\x})\big) \cap (S_4^+ \cup S_4^-)} \\
    &\qquad\qquad\qquad+ \card{\big(\supp(\x) \cup \supp(\tilde{\x})\big) \cap S_4^c} + \card{S_4^+ \cup S_4^-}\\
    & \geq 2\, \card{S_4^+ \cup S_4^-} + \card{S_4^c}.
  \end{align*}
  (The last inequality follows because, by construction,
  $S_i^\pm\subseteq\supp(\x)\cup\supp(\tilde{\x})$ for all $i \in [4]$.)

  Observe that for $i \in S_3^+ \cup S_4^+$, necessarily
  $\tilde{x}_i < 0$. Similarly, if $i \in S_3^- \cup S_4^-$, then
  $\tilde{x}_i > 0$. This shows that $\card{S_4^+} + \card{S_4^-} +
  \card{S_3^+} + \card{S_3^-} \leq \card{\supp(\tilde{\x})} \leq s$.

  Moreover, if $i \in S_2^+$ then $u_i > \delta^{\min}_i = -\ell_i$;
  furthermore, $-\tilde{x}_i \leq -\ell_i = \delta^{\min}_i$ (because
  $\tilde{\x}\in[\l,\u]_\Z$), which implies $x_i > 0$ (since
  $-\ell_i<z_i=x_i-\tilde{x}_i\leq x_i-\ell_i$). Similarly, if $i \in
  S_2^-$ then $\ell_i < - \delta^{\min}_i = -u_i$; thus, $-\tilde{x}_i
  \geq -u_i$, which shows that $x_i > 0$. Moreover, if $i \in S_4^+
  \cup S_4^-$ then $x_i \neq 0$. In total, this shows that
  $\card{S_4^+} + \card{S_4^-} + \card{S_2^+} + \card{S_2^-} \leq
  \card{\supp(\x)} \leq s$. Since all sets $S_i^\pm$ are disjoint or
  empty, this concludes the proof.
\end{proof}
\medskip

From the previous main theorem, we can derive the corresponding
characterizations for the remaining discrete sets:
\begin{corollary}\label{cor:recovery}\
  \begin{enumerate}[leftmargin=3ex]
  \item Let $X = [\0,\u]_\Z$. A matrix $\A \in \R^{m \times n}$ is
    $(s,[\0,\u]_\Z,0)$-good if and only if
    \begin{equation}\label{eq:P0NonnegBounded}
      \hspace*{-0.75em}\ker(\A) \cap \{\z \in [-\u,\u]_\Z \suchthat \card{\supp(\z;(\0,\u])}
      \leq s,\; \card{\supp(\z;[-\u,\0))} \leq s\} = \{\0\}.
    \end{equation}
  \item Let $X = \Z_+^n$. A matrix $\A \in \R^{m \times n}$ is
    $(s,\Z_+^n,0)$-good if and only if
    \begin{equation}\label{eq:P0Nonneg}
      \ker(\A) \cap \{\z \in \Z^n \suchthat \card{\supp(\z;\Z_{>0}^n)}
      \leq s,\; \card{\supp(\z;\Z_{<0}^n)} \leq s\} = \{\0\}.
    \end{equation}
  \item Let $X = [-\u,\u]_\Z$. A matrix $\A \in \R^{m \times n}$ is
    $(s,[-\u,\u]_\Z,0)$-good if and only if
    \begin{align}
      \nonumber\ker(\A) \cap & \{\z \in [-2 \cdot \u, 2 \cdot \u]_\Z \suchthat
      \; 2\, \card{\supp(\z;[-2 \cdot \u,-\u))}\\ 
      \label{eq:P0PlusMinus}  + & 2 \, \card{\supp(\z;(\u, 2 \cdot \u])} + \card{\supp(\z;[-\u,\0) \cup (\0,\u])} \leq 2s\} = \{\0\}.
    \end{align}
  \end{enumerate}
\end{corollary}
\begin{proof}\
  \begin{enumerate}
  \item We set $\l = \0$ in Theorem~\ref{thm:P0Master}. In this case,
    $\bdelta^{\min} = \0$ and $\bdelta^{\max} = \u$. Thus, only
    $S_2^+$ and $S_3^-$ can be nonempty, which yields
    \[
    C_\Z(\0,\u) \define \{\z \in [-\u,\u]_\Z \suchthat \card{S_3^-} \leq s,\;
    \card{S_2^+} \leq s,\; \card{S_3^-} + \card{S_2^+} \leq 2s\}.
    \]
    Since $\card{S_3^-} + \card{S_2^+} \leq 2s$ is redundant, $S_3^+ =
    \{i \suchthat z_i \in (0, u_i]\}$ and $S_2^- = \{i \suchthat z_i
    \in [-u_i,0)\}$, the condition from Theorem~\ref{thm:P0Master} is
    equivalent to that stated in~\eqref{eq:P0NonnegBounded}.
  \item This follows from the previous part by letting the components of
    $\u$ tend to infinity.
  \item We set $\l = -\u$ in Theorem~\ref{thm:P0Master}. In this case,
    $\bdelta^{\min} = \u$ and $\bdelta^{\max} = \u$. Thus, only
    $S_1^+$, $S_1^-$, $S_4^+$, and $S_4^-$ can be nonempty, which
    yields
    \begin{align*}
    C_\Z(-\u,\u) \define \{\z \in [-2 \cdot \u, 2 \cdot \u]_\Z \suchthat \card{S_4^-} +
    \card{S_4^+} \leq s,\; \card{S_4^-} + \card{S_4^+} \leq s&,\\
    2\,\big(\card{S_4^+} + \card{S_4^-}\big) + \card{S_1^-} + \card{S_1^+} \leq
    2s&\}.
    \end{align*}
    Since the first two constraints are identical and implied by the
    third one, we obtain from Theorem~\ref{thm:P0Master} the
    equivalent condition~\eqref{eq:P0PlusMinus}.\qedhere
  \end{enumerate}
\end{proof}
\medskip 

\begin{remark}\label{rem:binary}
  In particular, in the binary case (i.e., $X = [\0, \vones]_\Z$), we
  obtain that $A \in \R^{m \times n}$ is $(s,[\0,\vones]_\Z,0)$-good if and
  only if
  $\ker(\A) \cap \{\z \in \{-1,0,+1\}^n \suchthat \card{\{i \suchthat z_i =
    -1\}} \leq s,\; \card{\{i \suchthat z_i = 1\}} \leq s\} = \{\0\}$.
\end{remark}

\begin{remark}
  Note that Theorem~\ref{thm:Z-s-good} also follows from
  Theorem~\ref{thm:P0Master}: We let the components of $\l$ and $\u$
  simultaneously tend to $-\infty$ and $\infty$, respectively. Then
  $\delta^{\min}_i = -\ell_i \to\infty$ and $\delta^{\max}_i =  u_i \to \infty$ for
  all $i$. In this case, only $S_1^+$ and $S_1^-$ can be
  nonempty. Thus, the conditions of Theorem~\ref{thm:P0Master} reduce
  to $\card{S_1^+} + \card{S_1^-} \leq 2s$ and hence yield
  Theorem~\ref{thm:Z-s-good}.
\end{remark}

In the case of real-valued vectors, the proof of Theorem~\ref{thm:P0Master}
carries over directly and yields an analogous statement for
$\Prob{0}{[\l,\u]_\R}$ in which the vectors in $\ker(\A)$ are allowed to be
real (i.e., using the analogously defined $C_\R(\l,\u)$ instead of
$C_\Z(\l,\u)$). The same holds for results analogous to
Corollary~\ref{cor:recovery}; however, note that $\A$ is
$(s,[\0,\u]_\R,0)$-good if and only if it is $(s,\R^n_+,0)$-good, due to
the scalability of (real-valued) nullspace vectors. For clarity, we
summarize these new results for (bounded) real-valued sparse recovery in the
following Corollary.

\begin{corollary}\label{cor:P0Real}
   Let $\A \in \R^{m \times n}$.
   \begin{enumerate}
   \item Let $X = [\l,\u]_\R$ and define $\bdelta^{\min}$,
     $\bdelta^{\max}$, $S^{\pm}_i$ for $i \in [4]$ as well as $C_\R(\l,\u)$
     analogously to Theorem~\ref{thm:P0Master}. Then, $\A$ is
     $(s,[\l,\u]_\R,0)$-good if and only if
     $\ker(\A) \cap C_\R(\l,\u) = \{\0\}$.
   \item Let $X = \R_+^n$. Then, $\A$ is $(s,\R^n_+,0)$-good if and only if 
     \[
     \ker(\A) \cap \{\z \in \R^n \suchthat \card{\supp(\z;\R_{>0}^n)}
     \leq s,\; \card{\supp(\z;\R_{<0}^n)} \leq s\} = \{\0\}.
   \]
   \item Let $X = [\0,\u]_\R$. Then, $\A$ is $(s,[\0,\u]_\R,0)$-good if and only if
     it is $(s,\R_+^n,0)$-good.
   \item Let $X = [-\u,\u]_\R$. Then, $\A$ is $(s,[-\u,\u]_\R,0)$-good if and
     only if
     \begin{align*}
       \ker(\A) \cap & \{\z \in [-2 \cdot \u, 2 \cdot \u]_\R \suchthat
                       \; 2\, \card{\supp(\z;[-2 \cdot \u,-\u))}\\ 
       +\, &2\, \card{\supp(\z;(\u, 2 \cdot \u])} + \card{\supp(\z;[-\u,\0) \cup (\0,\u])} \leq 2s\} = \{\0\}.
     \end{align*}
   \end{enumerate}
\end{corollary}

\section{The $\ell_1$-case}\label{sec:l1case}

\noindent
As noted earlier, if it were not for the integrality constraints, the
problems $\Pone$,
$\PoneN$, $\Prob{1}{[-\u,\u]_\Z}$, $\PoneP$ and $\PonePM$ could all be
reformulated as linear programs (LPs). Hence, from the viewpoint of integer
programming, it is natural to ask under which conditions the LP relaxations
of these problems are guaranteed to have integral optimal solutions
themselves. To that end, we can resort to some well-established polyhedral
results often encountered in discrete and combinatorial optimization which
build on the concepts of (total) unimodularity and total dual
integrality. However, such general polyhedral integrality results do not
involve the aspect of solution sparsity (or uniqueness); we ne\-vertheless
provide several results obtained by this approach, but delegate this
discussion to Appendix~\ref{subsec:unimodTDIetc}.

A different viewpoint is taken by Keiper et al.~\cite{KeiKLP17}, who
consider recovery conditions and phase transitions, but restrict their
investigation to solutions in the sets $[\0,\vones]_\R$, $[-\vones,\vones]_\R$
or $[\0,2\cdot\vones]_\R$.

In contrast, we give complete characterizations of unique recoverability of
sparse integral vectors by $\ell_1$-minimization for all (general) cases,
based on extensions of the well-known nullspace property.  We begin with
uniform recovery guarantees in Section~\ref{subsec:l1nsps} and then provide
results for individual signal recovery in Section~\ref{subsec:P1indivRec}.

\subsection{Uniform Sparse Recovery Conditions}\label{subsec:l1nsps}

\noindent
To obtain succinct results for recovery of sparse integral vectors by
$\ell_1$-norm minimization, we turn to conditions on the nullspace of the
sensing matrix. The goal is to investigate the following property, similar
to Definition~\ref{def:0-good}:

\begin{definition}\label{def:1-good}
  Let $\A \in \R^{m \times n}$, $s \in [n]$, and $X \subseteq
  \R^n$.  The matrix $\A$ is \emph{$(s,X,1)$-good}, if \emph{every} $s$-sparse
  vector $\hat{\x} \in X$ is the unique solution of $\Prob{1}{X}$ with $\b
  = \A \hat{\x}$.
\end{definition}

In fact, if all $s$-sparse $\ell_1$-minimizers are unique, they also
solve the respective $\ell_0$-minimization problems:
\begin{proposition}\label{prop:1goodThen0good}
  If $\A$ is $(s,X,1)$-good, then it is $(s,X,0)$-good.
\end{proposition}
\begin{proof}
  Let $\A$ be $(s,X,1)$-good. Assume there is a minimizer $\z$ of
  $\Prob{0}{X}$ with $\b=\A\hat{\x}$ for some
  $s$-sparse~$\hat{\x}$. Then, $\A\z=\A\hat{\x}$ and
  $\norm{\z}_0\leq\norm{\hat{\x}}_0\leq s$, so that $\z=\hat{\x}$ must
  hold, since $\hat{\x}$ is (by definition of $(s,X,1)$-goodness) the
  unique minimizer of $\Prob{1}{X}$ with $\b=\A\hat{\x}$. Thus,
  $\card{S(s,X,\b)}=1$, i.e., $\A$ is $(s,X,0)$-good.
\end{proof}
\medskip 
For the sake of brevity, we will not explicitly repeat the
corresponding inferences regarding $(s,X,0)$-goodness in all the
following results pertaining to $(s,X,1)$-goodness, as they simply
follow from Proposition~\ref{prop:1goodThen0good}.  
\medskip

For $\A\in\R^{m\times n}$, a set $S \subseteq [n]$ and some $V
\subseteq \R^n$, we define the following two \emph{nullspace
  properties} (NSPs):
\begin{align*}
  \NSP{V}: && \norm{\v_S}_1 < \norm{\v_{S^c}}_1 & \qquad\forall\,\v\in (V\cap\ker(\A)) \setminus \{\0\},\\
  \NSPP{V}: && \v_{S^c} \geq 0 ~\Rightarrow~\vones^\top \v > 0 & \qquad\forall\,\v \in (V\cap\ker(\A)) \setminus \{\0\}.
\end{align*}
If a matrix $\A$ satisfies one of these conditions for \emph{all} sets
$S$ of cardinality $\card{S}\leq s$, we say that the respective NSP
\emph{of order $s$} is satisfied. 

In the continuous setting, nullspace properties are well-known to yield the
strongest results relating $\ell_1$-minimization to the recovery of sparse
vectors. For the sake of completeness, we summarize the fundamental such
results from the literature (rephrased in the notation of the present
paper) in the following theorem.

\begin{theorem}\label{thm:nspsReal}
  Let $\A\in\R^{m\times n}$ and $S \subseteq [n]$.
  \begin{enumerate}
  \item Every vector $\hat{\x} \in \R^n$ with $\supp(\hat{\x}) \subseteq S$
    is the unique solution of $\Prob{1}{\R^n}$ with
    $\b\define\A\hat{\x}$ if and only if $\A$ satisfies $\NSP{\R^n}$
    w.r.t.\ the set $S$. Moreover, $\A$ is $(s,\R^n,1)$-good if and only if $\A$
    satisfies $\NSP{\R^n}$ of order~$s$.
  \item Every vector $\hat{\x}\in\R^n_+$ with $\supp(\hat{\x})\subseteq S$
    is the unique solution of $\Prob{1}{\R^n_+}$ with $\b\define\A\hat{\x}$
    if and only if $\A$ satisfies $\NSPP{\R^n}$ w.r.t.\ the set $S$.
    Moreover, $\A$ is $(s,\R_+^n,1)$-good if and only if $\A$
    satisfies $\NSPP{\R^n}$ of order~$s$.
  \end{enumerate}
\end{theorem}
\begin{proof}
  For a proof of statement 1, see, e.g., \cite{FouR13}, and for statement 2,
  see~\cite{KhaDXH11}.
\end{proof}
\medskip

In fact, the proofs referenced for Theorem~\ref{thm:nspsReal} can almost
literally be translated to the case of $\Prob{1}{\Z^n}$ and
$\Prob{1}{\Z^n_+}$ by additionally requiring integrality of the nullspace
vectors. Thus, we immediately obtain the following novel result.

\begin{theorem}\label{thm:nspsZuZp}
  Let $\A \in \R^{m\times n}$ and $S \subseteq [n]$.
  \begin{enumerate}
  \item Every vector $\hat{\x}\in\Z^n$ with $\supp(\hat{\x})\subseteq S$
    is the unique solution of $\Prob{1}{\Z^n}$ with
    $\b\define\A\hat{\x}$ if and only if $\A$ satisfies $\NSP{\Z^n}$
    w.r.t.\ the set $S$.  Moreover, $\A$ is $(s,\Z^n,1)$-good if and only if $\A$
    satisfies $\NSP{\Z^n}$ of order~$s$.
  \item Every vector $\hat{\x}\in\Z^n_+$ with $\supp(\hat{\x})\subseteq S$
    is the unique solution of $\Prob{1}{\Z^n_+}$ with $\b\define\A\hat{\x}$
    if and only if $\A$ satisfies $\NSPP{\Z^n}$ w.r.t.\ the set $S$.
    Moreover, $\A$ is $(s,\Z_+^n,1)$-good if and only if $\A$
    satisfies $\NSPP{\Z^n}$ of order~$s$.
  \end{enumerate}
\end{theorem}

Similarly to Theorem~\ref{thm:Z-s-good}, for rational matrices there
is no difference between the standard (continuous) NSPs and their
integral counterparts, since rational kernel vectors can always be
rescaled to integrality:

\begin{corollary}\label{cor:rationalAnsps}
  Let $\A\in\Q^{m\times n}$. Then $\A$ satisfies $\NSP{\Z^n}$ if and only if
  it satisfies $\NSP{\R^n}$, and it satisfies $\NSPP{\Z^n}$ if and only if
  it satisfies $\NSPP{\R^n}$.
\end{corollary}

As a consequence, for rational data, signal integrality does not lead
to recoverability (by $\ell_1$-norm minimization) of lower sparsity
levels---i.e., larger number of nonzeros---than in the continuous
case. However, this situation again changes once the signal is
bounded.
\medskip

As a first criterion for $\PonePM$, we consider
NSP($[\l-\u,\u-\l]_\Z$), which leads to the following result.

\begin{proposition}\label{thm:boundedNSPsuff}
  If $\A\in\R^{m\times n}$ satisfies $\NSP{[\l-\u,\u-\l]_\Z}$ w.r.t.\
  a set $S\subseteq [n]$, then every vector $\hat{\x}\in[\l,\u]_\Z$
  with $\supp(\hat{\x})\subseteq S$ is the unique solution of
  $\PonePM$ with $\b\define\A\hat{\x}$ . Moreover, if $\A$ satisfies
  $\NSP{[\l-\u,\u-\l]_\Z}$ of order~$s$, then $\A$ is
  $(s,[\l,\u]_\Z,1)$-good.
\end{proposition}
\begin{proof}
  Assume $\A$ satisfies $\NSP{[\l-\u,\u-\l]_\Z}$ w.r.t.\ $S$. Suppose
  $\hat{\x}\in[\l,\u]_\Z$ has $\supp(\hat{\x})\subseteq S$, and let
  $\z\in[\l,\u]_\Z\setminus\{\hat{\x}\}$ satisfy
  $\A\z=\A\hat{\x}$. Then, $\v\define\hat{\x}-\z\in
  (\ker(\A)\cap[\l-\u,\u-\l]_\Z)\setminus\{\0\}$ and thus,
  \begin{align*}
    \norm{\hat{\x}}_1 &\leq \norm{\hat{\x}-\z_S}_1+\norm{\z_S}_1 = \norm{\v_S}_1+\norm{\z_S}_1\\
    &< \norm{\v_{S^c}}_1+\norm{\z_S}_1 = \norm{\z_{S^c}}_1+\norm{\z_S}_1=\norm{\z}_1.
  \end{align*}
  It follows that $\hat{\x}$ is the unique optimal solution of
  $\PonePM$ with $\b\define\A\hat{\x}$. Furthermore, by letting the
  set~$S$ vary, we immediately obtain the claim about uniqueness of
  all $s$-sparse solutions of $\PonePM$, i.e., $(s,X,1)$-goodness
  of~$\A$.
\end{proof}
\medskip

The above proof is a straightforward adaptation of the sufficiency
proof of the original results for $\Prob{1}{\R^n}$ and
$\NSP{\R^n}$ to the bounded-integers setting.  Unfortunately, the
condition of Proposition~\ref{thm:boundedNSPsuff} (i.e.,
$\NSP{[\l-\u,\u-\l]_\Z}$) is no longer necessary in the present case,
as the following toy example shows:

\begin{example}
  Let $\A=(1,2)$, $-\l=\u=\vones$ and $S=\{1\}$. Clearly, every vector
  in $[\l,\u]_\Z$ supported on $S$ (i.e., either $(0,0)^\top$,
  $(1,0)^\top$ or $(-1,0)^\top$) is the unique minimizer of $\PonePM$
  with the associated~$\b$. However, it holds that
  $(\ker(\A)\cap[-2\cdot\vones,2\cdot\vones]_\Z) \setminus\{\0\} =
  \{(-2,1)^\top,(2,-1)^\top\}$. Both of these vectors violate the
  condition of $\NSP{[\l-\u,\u-\l]_\Z}$, which here simply amounts to
  $\abs{v_1}<\abs{v_2}$ for all $\v=(v_1,v_2)^\top$ in the above
  nullspace subset.
\end{example}

We will give a complete characterization for sparse recovery via
$\PonePM$ later (see Theorem~\ref{thm:PonePMnsp} below), but first point out
a few more observations. The first one is again due to the scalability
of nullspace vectors:
\begin{corollary}\label{cor:luNSPRequivNSPR}
  For $\Prob{1}{[\l,\u]_\R}$ (with $\l\leq\0\leq\u$, $\l<\u$ both in
  $\R^n$), the analogous $\NSP{[\l-\u,\u-\l]_\R}$ is equivalent to the
  standard $\NSP{\R^n}$. 
\end{corollary}

Moreover, even though Proposition~\ref{thm:boundedNSPsuff} provides only a
sufficient condition for integral sparse recovery by means of
$\PonePM$, one can easily find examples which demonstrate that it is
already strictly weaker than its continuous analogon. Trivial examples
are obtained in cases in which there are no integral kernel vectors
satisfying the bounds $[\l-\u,\u-\l]_\Z$ other than $\0$ itself.  More
interestingly, consider the following case: 

\begin{example}\label{ex:boundedNSPsparsityDiff2}
  We revisit Example~\ref{ex:boundedNSPsparsityDiff1}: Let
  $-\l=\u=2\cdot\vones\in\Z^n$ with $n\geq 10$, let
  $\v=(2,-2,(\v^\prime)^\top)^\top$ and
  $\w=(\floor{\tfrac{n}{2}}, \floor{\tfrac{n}{2}}, (\w^\prime)^\top)^\top$
  with $\v^\prime\in\{-2,2\}^{n-2}$ and $\w^\prime\in\{-1,1\}^{n-2}$
  arbitrary, and let $\A$ be such that $\ker(\A) =
  \text{span}\{\v,\w\}$. (This means that up to elementary row operations,
  $\A$ has the form $(\I_{n-2},\V)$ with the identity matrix
  $\I_{n-2}\in\R^{(n-2)\times (n-2)}$ and
  $\V\define(\bar{\v},\bar{\w})\in\R^{(n-2)\times 2}$ for some
  $\bar{\v},\bar{\w}\in\R^{n-2}$ that can be obtained from $\v,\w$ by
  elementary row operations transforming $(\v,\w)$ into
  $(\V^\top,-\I_2)^\top$. Note also that $\V$ is rational, so $\A$ can even
  be chosen integral.) By construction, $\pm\v$ and $\pm 2\cdot\v$ are the
  only nonzero vectors in $\ker(\A) \cap [\l-\u,\u-\l]_\Z$. Moreover, for
  any $S\subset [n]$ with $\card{S}\leq s\define \floor{\tfrac{n}{2}} - 1$,
  it holds that
  $\norm{\pm\v_S}_1\leq 2s< 2\ceil{\frac{n}{2}} \leq \norm{\pm\v_{S^c}}_1$,
  and consequently also
  $\norm{\pm 2\cdot\v_{S}}_1<\norm{\pm 2\cdot\v_{S^c}}_1$, which means
  that~$\A$ satisfies $\NSP{[\l-\u,\u-\l]_\Z}$ of order~$s$. On the other
  hand, for $T=\{1,2\}$ we have
  $\norm{\w_T}_1 = 2 \cdot \floor{\tfrac{n}{2}} \geq n-1 > n-2 =
  \norm{\w_{T^c}}_1$, which reveals that $\A$ violates the
  $\NSP{[\l-\u,\u-\l]_\R}$---or equivalently, the standard continuous
  $\NSP{\R^n}$, cf. Cor.~\ref{cor:luNSPRequivNSPR}---for \emph{all} orders
  $t\geq 2$. (In other words, the standard continuous tools could only ever
  guarantee recovery of $1$-sparse vectors, which is of course trivial
  since then, $\b$ is just a rescaled column of $\A$.) In conclusion, the
  difference of recoverable sparsity orders $s-t$ can grow arbitrarily
  large with $n$---i.e., the example shows that integral basis pursuit with
  bounds is, in general, able to reconstruct integral signals up to much
  larger numbers of nonzeros than what could be guaranteed by
  integrality-oblivious results.
\end{example}

Furthermore, note that the previous example also shows that in the
presence of bounds, the integral and continuous nullspace properties
no longer coincide even for rational matrices $\A$.  

Finally, by setting $\l=-\u$, we obtain results analogous to
Propositon~\ref{thm:boundedNSPsuff} for the case $X=[-\u,\u]_\Z$:
\begin{corollary}\label{cor:boundedNSPsymm}
  If $\A\in\R^{m\times n}$ satisfies $\NSP{[-2 \cdot \u,2 \cdot
    \u]_\Z}$ w.r.t.\ a set $S\subseteq [n]$, then every vector
  $\hat{\x}\in[-\u,\u]_\Z$ with $\supp(\hat{\x})\subseteq S$ is the
  unique solution of $\Prob{1}{[-\u,\u]_\Z}$ with $\b\define\A\hat{\x}$ . Moreover,
  if $\A$ satisfies $\NSP{[-2 \cdot \u,2 \cdot \u]_\Z}$ of order~$s$,
  then $\A$ is $(s,[-\u,\u]_\Z,1)$-good.
\end{corollary}
\medskip

We now turn our attention to $\PoneP$. A sufficient condition can of course
be derived from Proposition~\ref{thm:boundedNSPsuff} again. We skip
the explict statement, since the following stronger result provides a full
characterization of recoverability for sparse nonnegative and upper-bounded
integral signals.

\begin{theorem}\label{thm:PonePnsp}
  Every vector $\hat{\x}\in[\0,\u]_\Z$ with $\supp(\hat{\x})\subseteq S$ is
  the unique optimal solution of $\PoneP$ with $\b\define\A\hat{\x}$ if and
  only if $\A\in\R^{m\times n}$ satisfies $\NSPP{[-\u,\u]_\Z}$ w.r.t.\
  $S$. Moreover, $\A$ is $(s,[\0,\u]_\Z,1)$-good if and only if~$\A$
  satisfies $\NSPP{[-\u,\u]_\Z}$ of order~$s$.
\end{theorem}
\begin{proof}
  We only prove the first statement, since the second one is again
  obtained immediately by letting the set $S$ vary. We modify the
  proof of Theorem~\ref{thm:nspsReal} part~2) (see~\cite{KhaDXH11}) to
  suit our setting: Suppose every $\hat{\x}\in[\0,\u]_\Z$ with
  $\supp(\hat{\x})\subseteq S \subseteq [n]$ is the unique minimizer
  of $\PoneP$ with $\b\define\A\hat{\x}$. Let
  $\0\neq\v\in\ker(\A)\cap[-\u,\u]_\Z$ and suppose
  $\v_{S^c}\geq\0$. Then,
  \begin{align*}
    \A\v=\0\quad\Leftrightarrow\quad&\A\v_{S}=\A(-\v_{S^c})\\
    \Leftrightarrow\quad&\A\v^+_S-\A\v^-_S=-\A\v_{S^c}\quad\Leftrightarrow\quad\A\v^-_S=\A(\v_{S^c}+\v^+_S),
  \end{align*}
  where $\v^\pm\define\max\,\{\0,\pm\v\}$ (component-wise), so that
  $\v=\v^+-\v^-$ and $\v^\pm\in[\0,\u]_\Z$. Obviously, $\v^-_S$ is
  supported on $S$ and $\v_{S^c}+\v^+_S\geq\0$. By construction,
  $\v^-_S$ uniquely solves $\PoneP$ with $\b_v\define\A\v^-_S$, so
  that $\norm{\v^-_S}_1<\norm{\v_{S^c}+\v^+_S}_1$. In fact, we obtain
  \begin{align*}
    &\norm{\v^-_S}_1<\norm{\v_{S^c}+\v^+_S}_1=\norm{\v_{S^c}}_1+\norm{\v^+_S}_1\\
    \Rightarrow\quad&\vones^\top\v = \norm{\v_{S^c}}_1+\norm{\v^+_S}_1-\norm{\v^-_S}_1 > 0.
  \end{align*}
  For the converse direction, suppose $\A$ satisfies
  $\NSPP{[-\u,\u]_\Z}$ w.r.t. a set $S\subseteq [n]$. Let
  $\hat{\x}\in[\0,\u]_\Z$ with $\supp(\hat{\x})\subseteq S$ and let
  $\z\in[\0,\u]_\Z\setminus\{\hat{\x}\}$ with
  $\A\z=\A\hat{\x}$. Consider $\v\define\z-\hat{\x}$; clearly, $\v\in
  (\ker(\A)\cap[-\u,\u]_\Z)\setminus\{\0\}$, and by construction, it
  holds that $\v_{S^c}=\z_{S^c}\geq\0$. By $\NSPP{[-\u,\u]_\Z}$,
  this implies
  \begin{align*}
    &0<\vones^\top\v=\norm{\v_{S^c}}_1+\norm{\v^+_S}_1-\norm{\v^-_S}_1\\
    \Leftrightarrow\quad &0<\norm{\z_{S^c}}_1+\norm{(\z-\hat{\x})^+_S}_1-\norm{(\z-\hat{\x})^-_S}_1=\norm{\z}_1-\norm{\hat{\x}}_1.
  \end{align*}
  To see the last equation, note that for $i\in S$, since
  $\hat{\x},\z\geq\0$, either $0\leq z_i-\hat{x}_i$ so that
  $\abs{(z_i-\hat{x}_i)^+}=\abs{z_i-\hat{x}_i}=z_i-\hat{x}_i=\abs{z_i}-\abs{\hat{x}_i}$
  and $\abs{(z_i-\hat{x}_i)^-}=0$, or $0<\hat{x}_i-z_i$ so that
  $\abs{(z_i-\hat{x}_i)^+}=0$ and
  $\abs{(z_i-\hat{x}_i)^-}=\abs{\hat{x}_i-z_i}=\hat{x}_i-z_i=\abs{\hat{x}_i}-\abs{z_i}$.
  Thus, $\norm{\hat{\x}}_1<\norm{\z}_1$, which concludes the proof.
\end{proof}
\medskip

For the continuous case, there is again no difference between the
$\nspop_+$ with bounds and the standard $\nspop_+$, since we can always
scale the kernel vectors accordingly:

\begin{corollary}
  For $\Prob{1}{[\0,\u]_\R}$, the analogous $\NSPP{[-\u,\u]_\R}$ is
  equivalent to the standard $\NSPP{\R^n}$.
\end{corollary}

As mentioned earlier, the result from Theorem~\ref{thm:PonePnsp} can be
transferred to the previously considered problem $\PonePM$ by utilizing a
standard variable split. We obtain the following recoverability
characterizations:
\smallskip

\begin{theorem}\label{thm:PonePMnsp}
  Let $\A\in\R^{m\times n}$ and $S \subseteq [n]$. Every vector
  $\hat{\x}\in[\l,\u]_\Z$ with $\supp(\hat{\x})\subseteq S$ is the
  unique optimal solution of $\PonePM$ with $\b\define\A\hat{\x}$ if
  and only if $(\A,-\A)$ satisfies
  $\NSPP{\left[\left(\begin{subarray}{r}-\u\\~\l\end{subarray}\right),\left(\begin{subarray}{r}~\u\\-\l\end{subarray}\right)\right]_\Z}$
  w.r.t.\ $S$.  Moreover, $\A$ is $(s,[\l,\u]_\Z,1)$-good if and only
  if $(\A,-\A)$ satisfies
  $\NSPP{\left[\left(\begin{subarray}{r}-\u\\~\l\end{subarray}\right),\left(\begin{subarray}{r}~\u\\-\l\end{subarray}\right)\right]_\Z}$
  of order~$s$.
\end{theorem}
\begin{proof}
  We split $\x=\x^+-\x^-$ with $\x^\pm\define\max\,\{\0,\pm\x\}$
  (component-wise). Thus, $\x^+\in[\0,\u]_\Z$ and $\x^-\in[\0,-\l]_\Z$
  when $\x\in[\l,\u]_\Z$. Then, we can rewrite $\PonePM$ as a problem
  in the form of nonnegative integral basis pursuit with upper bounds:
  \begin{equation}\label{eq:PonePMsplit}
    \min\;\left\{\left\lVert\left(\hspace*{-0.4em}\begin{array}{c}\x^+\\\x^-\end{array}\hspace*{-0.4em}\right)\right\rVert_1\suchthat(\A,-\A)\left(\hspace*{-0.4em}\begin{array}{c}\x^+\\\x^-\end{array}\hspace*{-0.4em}\right)=\b,~\left(\hspace*{-0.4em}\begin{array}{c}\x^+\\\x^-\end{array}\hspace*{-0.4em}\right)\in\left[\0,\left(\hspace*{-0.4em}\begin{array}{c}\u\\-\l\end{array}\hspace*{-0.4em}\right)\right]_{\Z}\right\}.
  \end{equation}
  Note that we may assume, without loss of generality, complementarity of
  $\x^+$ and $\x^-$, i.e., that $x^+_i \cdot x^-_i = 0$ for all $i$;
  otherwise, we could subtract $\min\,\{x^+_i,x^-_i\}$ from both values and
  thus reduce the objective, which can be written equivalently as
  $\vones^\top\x^++\vones^\top\x^-$. Hence, $\hat{\x}$ is the unique
  optimal solution of $\PonePM$ if and only if $\hat{\x}^+$ and
  $\hat{\x}^-$ form the unique minimizer of~\eqref{eq:PonePMsplit}.  The
  claims now follow from Theorem~\ref{thm:PonePnsp} applied to the
  reformulation~\eqref{eq:PonePMsplit} of~$\PonePM$.
\end{proof}
\medskip

\begin{remark}
  The condition from Theorem~\ref{thm:PonePMnsp} can be rephrased as
  follows: If $\supp(\hat{\x})\subseteq S$, then
  $((\hat{\x}^+)^\top,(\hat{\x}^-)^\top)^\top$ is supported on
  $T\define S_+\cup(n+S_-)\subseteq [2n]$, where
  $S_+\define\{i\in S:\hat{x}_i>0\}$ and $S_-\define\{i\in
  S:\hat{x}_i<0\}$. Since
  $\ker(\A,-\A)=\{\left(\begin{subarray}{c}\v\\\w\end{subarray}\right)\in\R^{2n}\suchthat\A\v=\A\w\}$
  and $T^c=S_+^c\cup(n+S_-^c)$, the variable split therefore yields
  that $(\A,-\A)$ satisfies
  $\NSPP{\left[\left(\begin{subarray}{r}-\u\\~\l\end{subarray}\right),\left(\begin{subarray}{r}~\u\\-\l\end{subarray}\right)\right]_\Z}$
  w.r.t.\ $S$ if and only if for all $\v\in[\0,\u]_\Z$ and $\w\in[\0,-\l]_\Z$ with $\A\v=\A\w$ and
  $\norm{\v}_0+\norm{\w}_0\geq 1$, the following implication holds true:
  \begin{equation*}
    \v_{S_+^c},\,\w_{S_-^c}\geq\0\quad\Rightarrow\quad\vones^\top(\v+\w)>0.
  \end{equation*}

  Note also that a similar condition could be derived for $\Pone$ by
  applying Theorem~\ref{thm:nspsZuZp} part 2) to the corresponding split
  formulation (which is of the form~$\PoneN$), but of course
  Theorem~\ref{thm:nspsZuZp} part 1) already provides a full (and simpler)
  characterization for sparse recovery by $\Pone$.
\end{remark}

\begin{corollary}\label{cor:PonePMnsp}
  Let $\A\in\R^{m\times n}$ and $S \subseteq [n]$.  Every vector
  $\hat{\x}\in[-\u,\u]_\Z$ with $\supp(\hat{\x})\subseteq S$ is the
  unique optimal solution of $\PonePM$ with $\b\define\A\hat{\x}$ if
  and only if $(\A,-\A)$ satisfies
  $\NSPP{\left[-\left(\begin{subarray}{r}\u\\\u\end{subarray}\right),\left(\begin{subarray}{r}\u\\\u\end{subarray}\right)\right]_\Z}$
  w.r.t.\ $S$.  Moreover, $\A$ is $(s,[-\u,\u]_\Z,1)$-good if and only
  if $(\A,-\A)$ satisfies
  $\NSPP{\left[-\left(\begin{subarray}{r}\u\\\u\end{subarray}\right),\left(\begin{subarray}{r}\u\\\u\end{subarray}\right)\right]_\Z}$
  of order~$s$.
\end{corollary}
\begin{proof}
  Set $\l=-\u$ and apply Theorem~\ref{thm:PonePMnsp}.
\end{proof}
\medskip

\begin{figure}[tb]
  \begin{center}
  \begin{tikzpicture}[scale=0.75, transform shape, >=stealth]
    \tikzstyle{Textknoten}=[draw=black!100,thick,inner sep=6pt, outer sep=4pt, minimum width = 15pt]
    \tikzstyle{Bogen}=[->,draw=black!100,thick]
    \tikzstyle{dBogen}=[<->,draw=black!100,thick]
    \node[Textknoten,fill=black!7, anchor=west,align=left] (nsp) at (-5,6) {$\NSP{\Z^n}$};
    \node[Textknoten,fill=black!7, anchor=west, align=left] (bnsp) at (-5,4.5) {$\NSP{[\l-\u,\u-\l]_\Z}$};
    \node[Textknoten,fill=black!7, anchor=west, anchor=west, align=left] (sbnsp) at (-4,3)
    {$\NSPP{\left[\left(\begin{subarray}{r}-\u\\~\l\end{subarray}\right),\left(\begin{subarray}{r}~\u\\-\l\end{subarray}\right)\right]_\Z}$ (for $(\A,-\A)$)};
    \node[Textknoten,fill=black!7, anchor=west, align=left] (nnsp) at (-5,1.5) {$\NSPP{\Z^n}$};
    \node[Textknoten,fill=black!7, anchor=west, anchor=west, align=left] (bnnsp) at (-5,0) {$\NSPP{[-\u,\u]_\Z}$};
    
    \node[Textknoten, rounded corners, anchor=east, align=left] (pz) at (10,6) {uniform recovery for $\Pone$};
    \node[Textknoten, rounded corners, anchor=east, align=left] (bpz) at (10,4.5) {uniform recovery for $\PonePM$};
    \node[Textknoten, rounded corners, anchor=east, align=left] (npz) at (10,1.5) {uniform recovery for $\PoneN$};
    \node[Textknoten, rounded corners, anchor=east, align=left] (bnpz) at (10,0) {uniform recovery for $\PoneP$};
    
    \draw[dBogen,double] (nsp) -- (pz);
    \draw[Bogen,double] (bnsp) -- (bpz);
    \draw[dBogen,double] (sbnsp.east) to [bend right=15] (bpz.south); 
    \draw[dBogen,double] (nnsp) -- (npz);
    \draw[dBogen,double] (bnnsp) -- (bnpz);
    \draw[Bogen,double] (nsp.west) to [bend right=35] (bnsp.west);
    \draw[Bogen,double] ([yshift=-0.2cm]nnsp.west) to [bend right=35] (bnnsp.west);
    \draw[Bogen,double] (nsp.west) to [bend right=35] (nnsp.west);
  \end{tikzpicture}
  \end{center}
  \caption{NSP-based recovery conditions for $\Prob{1}{X}$ for
    different $X\subseteq\Z^n$ and their relationships to each
    other. Arrows correspond to implications, whereas directions that
    are not depicted do not hold in general. The shorthand ``uniform
    recovery'' refers to guaranteed reco\-very of all vectors with a
    specific support $S$ (by NSPs w.r.t. $S$) and also to that of all
    $s$-sparse vectors (by NSPs of order $s$). Results pertaining to
    $\Prob{1}{[-\u,\u]_\Z}$ are not shown since these are simple
    special cases of those for $\Prob{1}{[\l,\u]_\Z}$.}
  \label{fig:NSPoverview}
\end{figure}
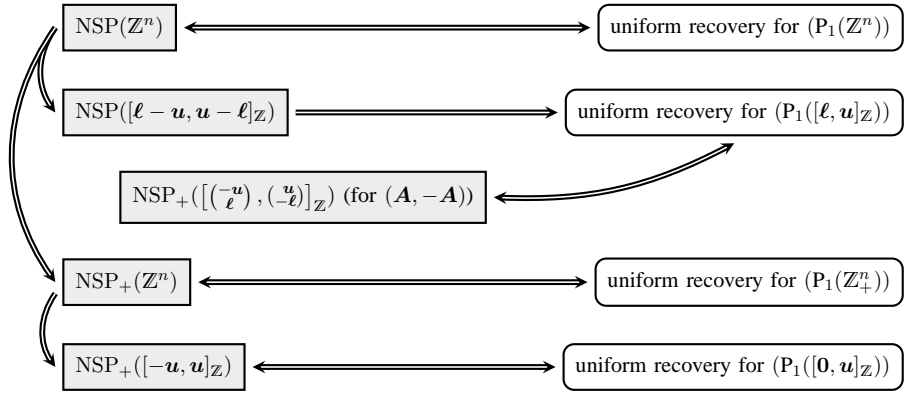

Naturally, the NSPs for larger integral sets imply those for smaller
sets. It is not hard to find examples that show that the converse
directions are false in general; for brevity, we do not list such
examples here, but provide an overview diagram to summarize our
results and display the implications, see
Figure~\ref{fig:NSPoverview}.

\subsection{Recovery of Individual Vectors}\label{subsec:P1indivRec}

\noindent
The focus so far was on conditions that guarantee the recovery of
\emph{all} vectors with a certain support or given sparsity level. In this
section, we consider similar conditions for the recovery of \emph{individual}
integral signals.

In the continuous setting, when $\Prob{1}{X}$ with $X\subseteq\R^n$
can be rewritten as a linear program, there are well-known
characterizations for recoverability of a specific vector $\hat{\x}$
as the unique $\ell_1$-minimizer, see, e.g.,
\cite[Theorem~4.26]{FouR13}. Attempting to directly transfer these
results to $\Pone$ gives the following sufficient (but not necessary) condition.

\begin{proposition}\label{thm:indivRecPone}
  A vector $\hat{\x}\in\Z^n$ with $\supp(\hat{\x})\subseteq
  S\subseteq [n]$ is the unique optimal solution of $\Pone$ with $\b\define\A\hat{\x}$
  if for all $\v\in (\ker(\A)\cap\Z^n) \setminus\{\0\}$, it holds that 
  \begin{equation}\label{eq:indivRecPone}
    \Big\lvert \sum_{i\in S}\sign(\hat{x}_i)\,v_i \Big\rvert < \norm{\v_{S^c}}_1.
  \end{equation}
\end{proposition}

The proof is completely analogous to (the sufficiency part in) that of
the above-cited theorem from~\cite{FouR13} and therefore omitted for
the sake of brevity.

Clearly, the condition from Proposition~\ref{thm:indivRecPone} is implied
by $\NSP{\Z^n}$, since $\norm{\v_S}_1\geq\abs{\sum_{i\in S} \sign(\hat{x}_i)\, v_i}$. 
Furthermore, the result also shows that it still makes no difference
whether we require~\eqref{eq:indivRecPone} to hold for all integral or all
rational vectors in the kernel of
$\A$ (the inequality is obviously scalable
by~$\alpha\in\N$), i.e., for $\A\in\Q^{m\times
  n}$ the condition is equivalent to its continuous analogon. However, the
condition loses necessity in the integral setting and indeed, it is not
hard to construct a simple counterexample.

For nonnegative vectors, a simple characterization of unique
recoverability is given next.

\begin{theorem}\label{thm:indivRecPoneN}
  A vector $\hat{\x}\in\Z^n_+$ is the unique optimal solution of
  $\PoneN$ with $\b\define\A\hat{\x}$ if and only if for all
  $\v\in (\ker(\A)\cap\Z^n) \setminus\{\0\}$, the following implication holds:
  \[
  \v+\hat{\x}\geq\0\quad\Rightarrow\quad\vones^\top\v>0.
  \]
\end{theorem}
\begin{proof}
  Since any other feasible solution can be written as the sum of
  $\hat{\x}$ and an integral nullspace vector~$\v$, $\hat{\x}$ is the
  unique point with smallest $\ell_1$-norm if and only if the
  objective contribution of every such nullspace vector is strictly
  positive.
\end{proof}
\medskip

Note that $\NSPP{\Z^n}$ implies the condition from
Theorem~\ref{thm:indivRecPoneN}, since for $\hat{\x}$ supported on
$S$, $\v+\hat{\x}\geq\0$ yields $\v_{S^c}\geq\0$ and $\NSPP{\Z^n}$ implies
that $\vones\T \v > 0$.

Finally, the two previous results can be extended directly to the
remaining cases $\PoneP$, $\Prob{1}{[-\u,\u]_\Z}$ and $\PonePM$, respectively. We omit the
completely analogous proofs.

\begin{proposition}\label{thm:indivRecPonePM}
  A vector $\hat{\x}\in[\l,\u]_\Z$ with $\supp(\hat{\x}) \subseteq
  S \subseteq [n]$ is the unique optimal solution of $\PonePM$ with $\b\define\A\hat{\x}$
  if for all vectors $\v\in (\ker(\A)\cap[\l-\u,\u-\l]_\Z)\setminus\{\0\}$, it holds that 
  \[
  \Big\lvert \sum_{i\in S}\sign(\hat{x}_i)\, v_i \Big\rvert < \norm{\v_{S^c}}_1.
  \]
\end{proposition}

\begin{corollary}\label{cor:indivRecPonePMsymm}
  A vector $\hat{\x}\in[-\u,\u]_\Z$ with $\supp(\hat{\x}) \subseteq S
  \subseteq [n]$ is the unique optimal solution of
  $\Prob{1}{[-\u,\u]_\Z}$ with $\b\define\A\hat{\x}$ if for all vectors $\v\in
  (\ker(\A)\cap[-2 \cdot\u,2\cdot\u]_\Z)\setminus\{\0\}$, it holds that
  \[
  \Big\lvert \sum_{i\in S}\sign(\hat{x}_i)\, v_i \Big\rvert < \norm{\v_{S^c}}_1.
  \]
\end{corollary}

\begin{theorem}\label{thm:indivRecPoneP}
  A vector $\hat{\x}\in[\0,\u]_\Z$ is the unique optimal solution of problem
  $\PoneP$ with $\b\define\A\hat{\x}$ if and only if for all
  $\v \in (\ker(\A)\cap[-\u,\u]_\Z) \setminus\{\0\}$, the following implication holds:
  \[
  \v+\hat{\x}\in[\0,\u]_\Z\quad\Rightarrow\quad\vones^\top\v>0.
  \]
\end{theorem}

The conditions for $\PonePM$ and $\Prob{1}{[-\u,\u]_\Z}$ are again only
sufficient, whereas that for $\PoneP$ gives a characterization of solution
uniqueness. It is worth mentioning that the conditions from
Proposition~\ref{thm:indivRecPonePM} (and
Corollary~\ref{cor:indivRecPonePMsymm}) and Theorem~\ref{thm:indivRecPoneP}
are strictly weaker than those from Proposition~\ref{thm:indivRecPone}
and Theorem~\ref{thm:indivRecPoneN}, respectively, as can easily be validated by
finding toy examples such as the one below confirming that ``bounds
on the variables matter''.
\begin{example}\label{ex:thm44}
Indeed, for instance, consider the matrices
\[
\A=\left(\begin{array}{rrrrrrr}
           3&0&0&0&3&-1&2\\
           0&3&0&0&0&1&1\\
           0&0&3&0&0&1&1\\
           0&0&0&3&-3&0&0
          \end{array}\right),\qquad
\V=\left(\begin{array}{rrr}1&-1&2\\0&1&1\\0&1&1\\-1&0&0\\-1&0&0\\0&-3&0\\0&0&-3\end{array}\right),
\]
where the columns of $\V$ span the nullspace of $\A$, and take
$\u=\vones$. Then, $(\cN(\A)\cap\Z^n)\setminus\{\0\}$ contains all
integer-valued linear combinations of the columns (say, $\v_1$, $\v_2$,
$\v_3$) of $\V$, whereas $(\cN(\A)\cap[-\u,\u]_\Z)\setminus\{\0\}$ contains
only $\pm \v_1$. For the vector $\hat{\x}=(1,1,1,0,0,0,0)^\top$, we can
thus easily verify via Theorem~\ref{thm:indivRecPoneP} that it is the
unique optimal solution for $\PoneP$:
\[
  \hat{\x}+\v_1\not\geq \0,\text{ but }\hat{\x}-\v_1\geq\0\text{ and
    indeed, }\vones^\top(-\v_1)=-1+1+1=1>0.
\]
With respect to $\PoneN$, however, the corresponding recovery condition
from Theorem~\ref{thm:indivRecPoneN} does not hold: Although
$\v^\prime\define-\tfrac{1}{3}\v_2-\tfrac{2}{3}\v_3\in\cN(\A)\cap\Z^n$ and
$\hat{\x}+\v^\prime\geq\0$, $\vones^\top\v^\prime = 0$, which shows that
here, $\hat{\x}$ is \emph{not} uniquely recoverable. Indeed, the point
$\hat{\x}+\v^\prime=(0,0,0,0,0,1,2)^\top\in\Z_+^n$ is another optimal
solution of $\PoneN$.
\end{example}

Finally, by employing the usual variable split, unique recoverability with
respect to $\PonePM$ and $\Prob{1}{[-\u,\u]_\Z}$ can also be fully
characterized: (Since we have seen all the arguments before, we skip the
proofs for brevity.)

\begin{theorem}\label{thm:charIndivRecPonePM}
  An $\hat{\x}\in[\l,\u]_\Z$ is the unique optimal solution of
  $\PonePM$ with $\b\define\A\hat{\x}$ if and only if for all
  $\v \in \left(\ker(\A,-\A)\cap\left[\left(\begin{subarray}{r}-\u\\~\l\end{subarray}\right),\left(\begin{subarray}{r}~\u\\-\l\end{subarray}\right)\right]_\Z\right) \setminus\{\0\}$, the following implication holds:
  \[
  \v+\left(\begin{subarray}{r}\hat{\x}^+\\\hat{\x}^-\end{subarray}\right)\in\left[\0,\left(\begin{subarray}{r}~\u\\-\l\end{subarray}\right)\right]_\Z\quad\Rightarrow\quad\vones^\top\v>0.
  \]
\end{theorem}

\begin{corollary}\label{cor:charIndivRecPonePMsymm}
  A vector $\hat{\x}\in[-\u,\u]_\Z$ is the unique optimal solution of
  problem $\Prob{1}{[-\u,\u]_\Z}$ with $\b\define\A\hat{\x}$ if and
  only if the following implication holds for all $\v \in
  \left(\ker(\A,-\A)\cap\left[-\left(\begin{subarray}{r}\u\\\u\end{subarray}\right),\left(\begin{subarray}{r}\u\\\u\end{subarray}\right)\right]_\Z\right)
  \setminus\{\0\}$:
  \[
  \v+\left(\begin{subarray}{r}\hat{\x}^+\\\hat{\x}^-\end{subarray}\right)\in\left[\0,\left(\begin{subarray}{r}\u\\\u\end{subarray}\right)\right]_\Z\quad\Rightarrow\quad\vones^\top\v>0.
  \]
\end{corollary}

Note that, since $\hat{\x}=\hat{\x}^+-\hat{\x}^-$, the condition in
Theorem~\ref{thm:charIndivRecPonePM} can be expressed equivalently as:
For all $\v\in\ker(\A)\cap[-\u,\u]_\Z$, $\w\in\ker(\A)\cap[\l,-\l]_\Z$
with $\norm{\v}_0+\norm{\w}_0\geq 1$, it holds that
$\v-\w+\hat{\x}\in[\l,\u]_\Z$ implies $\vones\T(\v+\w)>0$. An
analogous reformulation is, of course, also possible for the condition
in Corollary~\ref{cor:charIndivRecPonePMsymm}.

\section{Numerical Experiments}\label{sec:experiments}

\noindent
In this section, we present some computational experiments with the
recovery of integer signals as a proof-of-concept. The (mixed-)integer
problems were solved using Gurobi 7.5.2 on a linux cluster with Intel
Xeon E5-1620 quad core CPUs with 3.5 GHz, 10 MB cache size, and 32 GB
main memory. (For a primer on LP-based branch-and-bound, see,
e.g.,~\cite{Sch86}.)

\subsection{Solving the $\ell_0$-Problem for Binary Signals}\label{sec:experiments:binary}

\noindent
We begin with the case of binary signals, i.e., $X = \{0,1\}^n$. In this
case, $\Prob{0}{X}$ equals $\Prob{1}{X}$ and can be written
as
\begin{equation}\label{eq:P0binary}
  \min\; \{ \vones\T \x \suchthat \A \x = \b,\; \x \in \{0,1\}^n\}.
\end{equation}

In the first experiment, we deal with the solution
of~\eqref{eq:P0binary}. We generated a $64 \times 256$ matrix~$\A$
with random entries from $\{0, \dots, 99\}$. (Note that here, we avoid
the situation that, for some families of random matrices, all binary
signals can be reconstructed by solving~$\Prob{1}{[0,1]^n}$, i.e., the
LP relaxation of~\eqref{eq:P0binary}, if the number of measurements
satisfies $m\geq n/2$, cf.~\cite{Sto10}.) Then for $s = 8, 16, \dots,
256$ we generated a $\{0,1\}$-vector $\x^s$ with 1s at $s$ random
positions. The right hand side is then $\b^s \define \A \x^s$.

\begin{table}[t]
  \caption{Solving $\Prob{0}{\{0,1\}^n}$ via~\eqref{eq:P0binaryTrans} for a random $64 \times 256$
    matrix with entries in $\{0, \dots, 99\}$ and vector $\b = \A
    \tilde{\x}$, where the random vector~$\tilde{\x} \in \{0,1\}^n$ has~$s$
    nonzeros. (BKZ times were obtained on an Intel i7-3770 CPU with 3.4 GHz.)}
  \label{tab:prob}
  \vspace*{-0.75em}
  \begin{center}
    \footnotesize
    \begin{tabular}{@{}rrrrr@{\qquad}|@{\qquad}rrrrr@{}}\toprule
          &         &       & solver & BKZ      &       &         &       & solver & BKZ\\ 
      $s$ & opt. & nodes &time [s]& time [s] &   $s$ & opt. & nodes &time [s]& time [s] \\\midrule
      8   &     8   &     1 &    0.5 & 17.94 &      136 &   136   &  1335 &   16.2 & 18.16 \\
      16  &    16   &     1 &    0.5 & 19.05 &      144 &   144   &  1451 &   16.1 & 18.83 \\
      24  &    24   &     1 &    0.6 & 18.88 &      152 &   152   &  1219 &   12.1 & 18.28 \\
      32  &    32   &     1 &    1.1 & 18.13 &      160 &   160   &  1400 &   14.7 & 18.35 \\
      40  &    40   &    57 &    4.5 & 17.96 &      168 &   168   &  3309 &   23.8 & 18.13 \\
      48  &    48   &    47 &    4.0 & 18.09 &      176 &   176   &  1178 &   13.0 & 18.60 \\
      56  &    56   &   140 &    4.7 & 18.57 &      184 &   184   &  1614 &   15.1 & 18.48 \\
      64  &    64   &  1089 &    8.2 & 18.44 &      192 &   192   &  2229 &   15.9 & 18.02 \\
      72  &    72   &  1349 &   15.8 & 18.64 &      200 &   200   &   210 &    4.5 & 18.27 \\
      80  &    80   &  1799 &   21.2 & 18.40 &      208 &   208   &   132 &    3.9 & 18.07 \\
      88  &    88   &  2951 &   24.7 & 18.62 &      216 &   216   &    39 &    3.1 & 18.84 \\
      96  &    96   &  1408 &   16.9 & 18.47 &      224 &   224   &    50 &    3.3 & 18.50 \\
      104 &   104   &  1983 &   20.6 & 17.97 &      232 &   232   &     1 &    0.6 & 18.56 \\
      112 &   112   &  2050 &   24.1 & 18.55 &      240 &   240   &     1 &    0.6 & 18.60 \\
      120 &   120   & 16692 &  132.3 & 18.33 &      248 &   248   &     1 &    0.5 & 18.69 \\
      128 &   128   &  2034 &   21.7 & 18.51 &      256 &   256   &     1 &    0.5 & 18.33 \\
      \bottomrule
    \end{tabular}
  \end{center}
\end{table}

It turns out that directly solving problem~\eqref{eq:P0binary} is quite
hard -- only 12 instances can be solved to optimality within four hours; for
the solved instances the optimal solution was found right away and
optimality is proved fast. The remaining instances seem to be hopeless to
solve. Indeed, it is known that the so-called market-split instances, which
have a quite similar structure, are very challenging, see Cornu{\'e}jols
and Dawande~\cite{CorD99}. In fact, for such instances it has been known to
be hard to find a feasible solution or decide that none exists in practice
using standard solution techniques. Then, Aardal et al.~\cite{AarBHLS00}
observed that using basis reduction techniques, the market-split instances
can be transformed such that they can be solved easily for medium-sized
instances. We tested this transformation, but it turned out to be
inefficient, possibly because, to retain the objective function, one
needs to keep the original variables. Even branching on the transformed
variables first does not help here. However, we also tested the so-called
rangespace formulation of Krishnamoorthy and Pataki~\cite{KriP09}. Here, we
computed a unimodular matrix~$\U \in \Z^{n \times n}$ such that the columns
of $\left(\begin{subarray}{c}\A\\[0.1em]\I\end{subarray}\right)\U$ are almost orthogonal and of similar length. The
resulting model is then
\begin{equation}\label{eq:P0binaryTrans}
\min\; \{\vones\T \U \x \suchthat \A \U \x = \b,\; \0 \leq \U \x \leq \vones\}.
\end{equation}
The intuition is that the corresponding polytope is transformed to be more
``round''. This makes it easier for variable-branching based
branch-and-bound solvers to find feasible solutions and then prove
optimality.

The results for solving~\eqref{eq:P0binaryTrans} are shown in
Table~\ref{tab:prob}. The columns provide the sparsity level~$s$, the
optimal value of~\eqref{eq:P0binaryTrans} and therefore
of~\eqref{eq:P0binary}, the number of nodes in the branch-and-bound tree,
and the running time in seconds. We use the library
\texttt{fplll}~\cite{fplll} and its python interface
\texttt{fpylll}~\cite{fpylll} using the Block Korkin-Zolotarev (BKZ) basis
reduction technique, see Schnorr~\cite{Sch87}. The corresponding time in
seconds is given in the last column.

The results show that the optimal solution always coincides
with the sparsity level~$s$ used to construct the instances. Moreover, all
instances can be solved quite fast after performing basis reduction. We
have to note, however, that this approach will not scale well, since the
basis reduction algorithms will take a significant time for larger
instances and the solution of the instances as well. Nevertheless our
results hopefully motivate research to improve the presented techniques.

The general behavior of the solving times is quite typical for
sparsity-related problems. They tend to be small for small sparsity
levels~$s$ or $s$ close to the number of variables. One explanation is that
the number of 0/1 solutions satisfying the sparsity bound with equality is small, namely
$\binom{n}{s}$.

The results in Table~\ref{tab:prob} also suggest that the vectors~$\x^s$
used for the construction of the instances are in fact the only feasible
integer points. This can be tested by adding the constraint
\[
\sum_{i \in [n]:\; x^s_i = 1} (1 - x_i) + \sum_{i \in [n]:\; x^s_i = 0} x_i \geq 1.
\]
to the model~\eqref{eq:P0binary}. If the problem turns out to be
infeasible, $\x^s$ is the unqiue solution. In fact, the time to prove infeasibility of all
these instances is less than 0.1 seconds. Note,
however, that $\x^s$ is usually not known.

Moreover, the relaxations of the instances contain different fractional
solutions, which can be seen by the fact that in general there is more than
one branch-and-bound node. The results also show that the fractional
solutions have smaller $\ell_1$-norm than the integral solution. Thus,
uniqueness is enforced by the integrality condition and cannot be recovered
by using $\ell_1$-minimization for $X = [0,1]^n$.  \smallskip

Finally, note that the solution performance also depends on the size
of the coefficients in the matrix~$\A$. If we use a random $64 \times
256$ binary matrix, i.e., with entries from $\{0,1\}$, the problems
become harder to solve. In this case, \eqref{eq:P0binaryTrans} can
only be solved for 12 instances within four hours. Nevertheless, the
optimal values agree with~$s$ in all these instances. Again for all
instances, uniqueness of~$\x^s$ is easily proven.

\subsection{Recoverability Test for Binary Signals}\label{sec:recovery:binary}

\noindent
As a next step, we check whether the recoverability test of
Remark~\ref{rem:binary} allows to guarantee unique solutions. For $X =
\{0,1\}^n$, this test can be modeled as
\begin{align}
  \nonumber \max\; \{ \vones\T \v + \vones\T \w \suchthat\; & \A \v - \A \w = \0,\; \vones\T
  \v \leq s,\; \vones\T \w \leq s,\\
\label{eq:Recover}  & v_i + w_i \leq 1 \;\forall\, i \in [n],\; \vones\T \v \geq \vones\T \w;\; \v,\; \w \in \{0,1\}^n\},
\end{align}
where the last inequality removes symmetry with respect to sign flips
(i.e., scaling by
$-1$). We model $\z\in\{0,\pm 1\}^n\cap\cN(\A)$ as $\z=\v-\w$ with
$\v$, $\w\in\{0,1\}^n$ and $\v-\w\in\cN(\A)$; the constraints $v_i+w_i\leq 1$
for all~$i$ assert that
$\vones\T \v\leq s$ iff $\abs{\{i\suchthat z_i=1\}}\leq s$ and
$\vones\T \w\leq s$ iff $\abs{\{i\suchthat z_i=-1\}}\leq s$. Thus,
this formulation yields that the matrix $\A$ is $(s,X,0)$-good if and only
if the optimal objective is 0. Then one can use a cutoff value and stop
the computation as soon as a solution with positive objective is found.

Alternatively, the following model can be used:
\begin{align}
\nonumber  \min\; \{ \vones\T \v + \vones\T \w \suchthat\; & \A \v - \A \w = \0,\; \vones\T
  \v \leq s,\; \vones\T \w \leq s,\; \vones\T(\v+\w)\geq 1,\\
  \label{eq:RecoverAlt}  &v_i + w_i \leq 1 \;\forall\, i \in [n],\; \vones\T\v\geq\vones\T\w;\; \v,\; \w \in \{0,1\}^n\}.
\end{align}
Here, $\A$ is $(s,X,0)$-good if and only if this problem is infeasible. In
practice, \eqref{eq:RecoverAlt} performed worse than~\eqref{eq:Recover}.

To illustrate the behavior of~\eqref{eq:Recover}, we first consider a
matrix $\A$ of size $32 \times 64$ with random entries from $\{0, \dots,
99\}$. We transformed the problem using basis reduction,
as described above. In this case, all instances are solved within a few
seconds and recoverability is proven.
We also performed a similar test using a random $32 \times 96$ matrix with
random entries from $\{0, \dots, 99\}$. However, only instances up to $s =
6$ could be solved using the transformed problem with basis reduction
within four hours; recoverability could be proven for each of these
cases. No instance could be solved for the original formulation.

\begin{table}[t]
  \caption{Results of the recoverability formulation~\eqref{eq:Recover} for a $32 \times 64$
    random binary matrix and binary signals.}
  \label{tab:recover:binary}
  \vspace*{-0.75em}
  \begin{center}
    \footnotesize
    \setlength{\tabcolsep}{0ex}
    \begin{tabular*}{\textwidth}{@{\extracolsep{\fill}}rrrrrrrrrr@{}}\toprule
      $s$ & 10 & 11 & 12 & 13 & 14 & 15 & 16 & 17 & 18\\
      best obj. & 0 & 0 & 0 & 0 & 0 & 0 & 0 & 33 & 33\\
      time [s] & 284.1 & 834.5 & 1014.6 & 8051.6 & 6270.9 & 11632.6 & 8610.2 & 14400.0 & 14400.0\\
      \bottomrule
    \end{tabular*}
  \end{center}
\end{table}

Not surprisingly, the recoverability behavior depends on the sizes of the
coefficients in the matrix~$\A$. If we consider a $32 \times 64$ binary
matrix, recoverability can be proven for $s \leq 16$. Starting from $s =
17$, the optimal value is positive, i.e., no universal recovery holds, see
Table~\ref{tab:recover:binary}.
For $32 \times 96$ matrices the picture is similar. One can prove
recoverability up to $s = 14$. In both cases, applying basis reduction does
not help to speed up the solution process.

\subsection{Continuous Signals in the Unit Interval}

\noindent
In a next step, we consider the relaxed version $\Prob{0}{[\0,\vones]_\R}$
and compare it to the integral problem $\Prob{0}{[\0,\vones]_\Z}$. The goal
is to quantify the effect of requiring the signals to be integer on
recovery guarantees.

In the first experiment for continuous signals, we consider the recovery
test of Corollary~\ref{cor:P0Real} specialized to $X = [\0,\vones]_\R$,
which can be modeled as
\begin{align}
  \nonumber\max\; \{ \vones\T (\v + \w) \colon & \A \v - \A \w = \0,\; \0 \leq \v \leq
  \y,\; \0 \leq \w \leq \z,\; y_i + z_i \leq 1 \;\forall\, i \in [n],\\
\label{eq:RecoverUnit}  & \vones\T \y \leq s,\; \vones\T \z \leq s,\; \vones\T \v \geq \vones\T \w,\; \y,\; \z \in \{0,1\}^n\}.
\end{align}

As for~\eqref{eq:Recover}, the matrix $\A$ is $(s,X,0)$-good if and only if
the optimal objective is~$0$. In fact, for the same $32 \times 64$ random
matrix with entries from $\{0, \dots, 99\}$ from above, formulation
\eqref{eq:RecoverUnit} yields a positive optimal value for every~$s
\in [64]$, i.e., $\Prob{0}{[0,\vones]_\R}$ is never universally unique.
For the $32 \times 64$ \emph{binary} matrix from above, it turns out that
for $s = 10, \dots, 16$ the instances cannot be solved within a time limit
of one hour, and the results for these sparsity levels remain
inconclusive. For $s = 17$ and $s = 18$, a solution with positive objective
could be found, as in the integral case (cf.\
Table~\ref{tab:recover:binary}).

Next, we directly consider $\Prob{0}{[\0,\vones]_\R}$, which can be
written as
\begin{equation}\label{eq:P0relax}
  \min\; \{ \vones\T \y \suchthat \A \x = \b,\; 0 \leq \x \leq \y,\; \y \in \{0,1\}^n\}.
\end{equation}
Note that solving~\eqref{eq:P0relax} is \NP-hard by the same arguments
as used to prove Proposition~\ref{prop:complexity}.

Using the same $64 \times 256$ matrix and instances as shown in
Table~\ref{tab:prob}, it is possible to solve the instances with $s =
8$, $16$, $24$, $32$, $40$, $48$ (and $s = 216$, $224$, $232$, $240$,
$248$, $256$) within four hours. In all these cases, the (integral)
solution $\tilde{\x}$ used to generate~$\b$ is recovered. This shows
that adding bounds on the variables can result in quite strong
(individual) recovery guarantees, even if no integrality requirements
are imposed. For the binary $64 \times 256$ matrix from above, one can
solve $s = 8$, $16$, $24$, $32$, $224$, $232$, $240$, $248$, $256$
with the same conclusions.

Summarizing, the computations in this section do not show a
difference in the recovery properties between $X = [\0,\vones]_\Z$ and
$X = [\0,\vones]_\R$. Nevertheless, the solution performance might be
different. Moreover, this property does not hold in general, as shown
by Example~\ref{ex:binaryIntVSCont} earlier---there, the given
solutions are also $\ell_0$-minimizers of the respective problems and
the constraint $\x\leq\vones$ can be added, since it is already implied
by the data and nonnegativity.

\subsection{Signals With Values $0$, $1$ and $2$}

\noindent
In a next experiment, we consider $\Prob{0}{X}$ and $\Prob{1}{X}$ for the
cases $X = [\0, 2 \cdot \vones]_\Z$ and $X = [\0, 2 \cdot \vones]_\R$. For
the latter, we work with a formulation similar to~\eqref{eq:P0relax}. We
use a binary matrix of size $24 \times 72$.

\begin{table}[t]
  \caption{Solving~$\Prob{0}{[\0, 2 \cdot \vones]_\Z}$, $\Prob{0}{[\0,
      2 \cdot \vones]_\R}$, $\Prob{1}{[\0, 2 \cdot \vones]_\Z}$, and
    $\Prob{1}{[\0, 2 \cdot \vones]_\R}$ for a random $24 \times 72$
    binary matrix and vectors $\b$ generated by vectors~$\tilde{\x}$ of support size $s$.}
  \label{tab:prob012}
  \vspace*{-0.75em}
  \footnotesize
  \begin{center}
    \begin{tabular}{@{}rrr@{\qquad}rr@{\qquad}rrr@{\qquad}rr@{}}\toprule
                & \multicolumn{2}{c}{$\Prob{0}{[\0, 2 \cdot \vones]_\Z}$}
                & \multicolumn{2}{c}{$\Prob{0}{[\0, 2 \cdot \vones]_\R}$}
                & \multicolumn{3}{c}{$\Prob{1}{[\0, 2 \cdot \vones]_\Z}$}
                & \multicolumn{2}{@{}c@{}}{$\Prob{1}{[\0, 2 \cdot \vones]_\R}$}\\
                \cmidrule(r){2-3}\cmidrule(r){4-5}\cmidrule(r){6-8}\cmidrule{9-10}
      $s$& $\norm{\cdot}_0$ & time [s] & $\norm{\cdot}_0$ & time [s] &
      $\norm{\cdot}_1$ & $\norm{\cdot}_0$ & time [s] & $\norm{\cdot}_1$ & $\norm{\cdot}_0$\\
      \midrule
      12 &        12 &     0.0 & 12 &     0.0 &  19 & 12 &  0.0 & 19.00 & 12 \\
      16 &        16 &     5.6 & 16 &    17.3 &  22 & 16 &  0.1 & 19.54 & 17 \\
      20 &        20 &    27.5 & 20 &     5.0 &  33 & 25 &  0.5 & 30.81 & 29 \\
      24 &        24 &   177.1 & 24 &   267.0 &  34 & 24 &  0.9 & 32.15 & 29 \\
      28 & $\leq$ 27 & 14000.0 & 27 &  5817.0 &  42 & 30 &  4.6 & 40.47 & 34 \\
      32 & $\leq$ 30 & 14000.0 & 29 &  2970.9 &  49 & 34 & 12.2 & 47.75 & 36 \\
      36 & $\leq$ 29 & 14000.0 & 28 &  5844.6 &  48 & 32 &  0.6 & 46.80 & 35 \\
      40 & $\leq$ 34 & 14000.0 & 33 &  1371.3 &  56 & 35 & 10.1 & 54.50 & 42 \\
      44 & $\leq$ 35 & 14000.0 & 34 & 12146.3 &  58 & 40 &  1.3 & 57.34 & 42 \\
      48 & $\leq$ 41 & 14000.0 & 39 &  4800.5 &  69 & 44 &  3.4 & 67.81 & 46 \\
      52 &        46 &  5118.2 & 45 &   448.0 &  81 & 53 & 12.5 & 79.02 & 52 \\
      56 & $\leq$ 44 & 14000.0 & 44 &  2443.7 &  79 & 48 &  0.2 & 77.53 & 51 \\
      60 & $\leq$ 48 & 14000.0 & 46 &  2846.0 &  83 & 51 &  3.3 & 82.04 & 55 \\
      64 & $\leq$ 51 & 14000.0 & 49 &   262.5 &  91 & 57 &  2.5 & 89.48 & 56 \\
      \bottomrule
    \end{tabular}
  \end{center}
\end{table}

The results are given in Table~\ref{tab:prob012}. It turns out that the
integer program $\Prob{0}{[\0, 2 \cdot \vones]_\Z}$ is quite hard to solve:
only five instances could be solved to optimality within four hours; using
a rangespace formulation similar to~\eqref{eq:P0binaryTrans} does not
improve the situation. One main difference to the computations in
Section~\ref{sec:experiments:binary} is that the generating
vectors~$\tilde{\x}$ are not necessarily recovered, starting from $s =
28$. Indeed, the value of $\Prob{0}{[\0, 2 \cdot \vones]_\Z}$ is often less than
$s$; note that the table provides the value of the best primal solution
found during the run time, if the problem could not be solved to
optimality.

The continuous problem $\Prob{0}{[\0, 2 \cdot \vones]_\R}$ recovers the
original solution up to $s = 24$. It is relatively hard to
solve, but easier than $\Prob{0}{[\0, 2 \cdot \vones]_\Z}$.

The results for using an $\ell_1$-objective are also shown in
Table~\ref{tab:prob012}. Problem $\Prob{1}{[\0, 2 \cdot \vones]_\Z}$ can be
solved quite fast, but produces solutions of different sparsity than
$\Prob{0}{[\0, 2 \cdot \vones]_\Z}$. In fact, for $s = 20$,
$\Prob{1}{[\0, 2 \cdot \vones]_\Z}$ fails to recover~$\tilde{\x}$,
while $\Prob{0}{[\0, 2 \cdot \vones]_\Z}$ and
$\Prob{0}{[\0, 2 \cdot \vones]_\R}$ are successful.

The continuous counterpart $\Prob{1}{[\0, 2 \cdot \vones]_\R}$ amounts to
the solution of one LP and is very fast; we therefore do not list running
times for this variant in Table~\ref{tab:prob012}. However, it never
recovers the generating solutions $\tilde{\x}$ (except for $s = 12$) and
produces significantly denser solutions, demonstrating the stronger
reconstructability properties using integer variables.

In conclusion, these instances have a similar behavior to classical
compressed sensing: Reconstruction is only possible up to a certain
sparsity level. Moreover, this setting shows that there is a trade-off
between sparsity and the integrality requirement in view of (near)
recovery.

\section{Concluding Remarks}\label{sec:coda}

\noindent
It should not come as a surprise that the integrality-aware recovery
conditions are similar to their continuous counterparts. The core
argument is that a given feasible $\x$ with sparsity
$s$ can be modified to a different feasible~$\hat{\x}$ with sparsity
$\hat{s}<s$ if and only if there exists a nullspace vector
$\v = \x - \hat{\x}$ satisfying further constraints
that ensure feasibility. Crucial differences
arise with respect to possibilities to scale $\v$, integrality of $\v$, and
constraining $\v$ such that $\x + \v$ obeys possible bounds. The scaling
aspect renders many conditions in the continuous case to be equivalent, but
is more restricted in the discrete case---at least if bounds are present.
Provided the matrix $\A$ is rational, in the $\ell_1$-case, the known
conditions from the continuous setting indeed also hold in the discrete
settings if the variables are unbounded or nonnegative
(Theorem~\ref{thm:nspsZuZp} and Corollary~\ref{cor:rationalAnsps}). In the
$\ell_0$-case, this is true for the unbounded case (Theorem~\ref{thm:Z-s-good}), but no longer for the
nonnegative case (Corollary~\ref{cor:recovery}). Moreover, if $\A$ is allowed to be irrational, the
conditions are no longer equivalent in the above cases
(Proposition~\ref{thm:IrrationalRecovery}). Thus, the results obtained in
this paper demonstrate that integrality truly makes a difference here. Naturally,
bounds have a big influence on the admissable nullspace vectors in all
recovery conditions, but less prominently in the continuous setting,
especially with $\ell_1$-objective (cf., e.g., Corollary~\ref{cor:P0Real}
vs. Corollary~\ref{cor:luNSPRequivNSPR} regarding (P$_0([\ell,u]_\R)$) and
(P$_1([\ell,u]_\R)$), respectively).

Nevertheless, various aspects of integral sparse recovery are yet
unexplored. For instance, while the results obtained in the present paper
pertain to quite fundamental problems, it is also very important to explore
the stability and robustness of the recovery problems if the measurements
are corrupted by noise. In the continuous case, many explicit bounds on
recovery errors are known (i.e., estimates on how far away from the sought
true signal the solution of a recovery problem may be), but it seems no
such investigations have so far been carried out assuming signal
integrality. Similarly, it is of interest to see how integrality
constraints influence (probabilistic) bounds on the minimum number of
measurements needed to ensure unique recoverability under certain matrix
conditions. Also, one could consider integrality in the context of the
so-called cosparse (analysis) model.

The practical solution of all associated optimization problems involving
integrality remains challenging, similar to the exact solution of
$\Prob{0}{\R^n}$, cf.\ \cite{JokP08}. Thus, to harvest the benefits of
improved recovery capabilities when signal integrality is known in
practice, the development of further heuristics or approximation schemes as
well as exact solution algorithms for sparse recovery problems with
integrality constraints remains a vitally important task. In particular, it
should be worth looking into combining the modern general-purpose integer
programming solvers with problem-specific components like cutting planes,
branching and domain propagation rules or heuristics in order to improve
practical performance. (While such solution approaches are beyond the scope of the present
paper, the very promising results from the recent work~\cite{Til19} on
computing $\spark(\A)$---also an \NP-hard task---indicate that such
dedicated solvers may indeed achieve significant gains compared to
black-box methods for ``sparsity problems'' arising in the context of
compressed sensing.)

Finally, the same need for algorithm development can be expressed regarding
the actual practical evaluation of sparse recovery conditions such as the
various NSPs. Indeed, the computational complexity status of most NSPs
encountered in the present paper is apparently still open (checking the
well-known one from the continuous setting, NSP($\R^n$), is
\NP-hard~\cite{TilP14}), as is the related question whether such NSP
evaluations may be easy for certain (nontrivial) special classes of
matrices.

\section*{Acknowledgments}

\noindent
The authors would like to thank the two anonymous referees for their
valuable comments, which helped to improve the paper. They also thank
Felipe Serrano for providing his basis reduction code. This work was
partially supported by the EXPRESS project within the DFG priority program
CoSIP (DFG-SPP 1798).

\begin{small}
\setlength{\bibsep}{0pt plus 0.3ex}
\bibliographystyle{elsarticle-num}
\bibliography{preprint_dam_integercs}
\end{small}

\begin{appendices}
\section{-- ~~Integral LP Relaxations}\label{subsec:unimodTDIetc}

\noindent
Focussing purely on the integrality aspect (i.e., without incorporating
sparsity and solution uniqueness considerations explicitly), the
relationship of $\ell_1$-minimization problems to
integer linear programming provides some insights. For terminology from
polyhedral theory, broad overviews and collections of classical results, we
refer to the books by Schrijver~\cite{Sch86} and Korte and
Vygen~\cite{KorV12}.

We begin by considering the following question: When does the LP
relaxation $\ProbLP{1}{X}$ of $\Prob{1}{X}$ have integral optimal
solutions for every right hand side vector $\b$?  A first answer can
be obtained for unimodular matrices $\A\in\Z^{m\times n}$, i.e., those
for which every regular $m\times m$ submatrix has determinant $\pm 1$:

\begin{proposition}\label{prop:unimodularA}
  Let $\A\in\Z^{m\times n}$ be unimodular with $\rank(\A)=m\leq n$ and
  let $X = \Z^n$ or $X = \Z^n_+$. Then, for every $\b \in \Z^m$, the
  LP relaxation $\ProbLP{1}{X}$ has an integral optimal (vertex) solution,
  i.e., it lies in $X$.
\end{proposition}
\begin{proof}
  First, we consider $X=\Z^n_+$ and the LP
  \begin{equation}\label{eq:P1pLP}
    \min\;\big\{\norm{\x}_1\suchthat \A\x=\b,~\x\geq\0\big\} = \min\;\big\{\vones^\top \x\suchthat \A\x=\b,~\x\geq\0\big\}.\tag{$\text{P}_{1}^{\text{LP}}(\Z^n_+)$}
  \end{equation}
  Standard results (cf., e.g., \cite{Sch86}) show that unimodularity
  of $\A$ is equivalent to the integrality of the polyhedron
  $\{\x:\A\x=\b,\,\x\geq\0\}$ for every $\b\in\Z^m$. In particular,
  there exists an integral optimal vertex solution
  for~\eqref{eq:P1pLP}, since~\eqref{eq:P1pLP} always has a finite
  value; this solution is also optimal for $\PoneN$.

  Now consider $X=\Z^n$. By means of the standard variable split
  $\x=\x^+-\x^-$ with $\x^+\define\max\,\{\0,\x\}$, $\x^-\define\max\,\{\0,-\x\}$
  (component-wise), we transform $\ProbLP{1}{\Z^n}$ into the LP
  \begin{equation}\label{eq:P1asLP}
    \min\;\big\{\vones^\top\x^++\vones^\top\x^-\suchthat \A\x^+-\A\x^-=\b,~\x^+\geq\0,~\x^-\geq\0\big\}.
  \end{equation}
  Clearly, if $\A$ is unimodular, then so is $(\A,-\A)$, and
  since~\eqref{eq:P1asLP} is of the same form as~\eqref{eq:P1pLP}, the
  conclusion carries over. It remains to note that every integral
  optimal vertex solution $(\bar{\x}^+,\bar{\x}^-)$
  of~\eqref{eq:P1asLP} yields a corresponding integral optimal
  solution $\bar{\x}\define\bar{\x}^+-\bar{\x}^-$ for $\ProbLP{1}{\Z^n}$,
  which also solves $\Prob{1}{\Z^n}$.
\end{proof}
\medskip

Strengthening the structural assumption on $\A$, we obtain analogous
results for the remaining cases of integral sets considered in this
paper:

\begin{proposition}\label{prop:tuA}
  Let $\A\in\Z^{m\times n}$ be totally unimodular (i.e., every square
  submatrix has determinant $0$ or $\pm 1$) with $\rank(\A) = m \leq n$ and
  let $X = \Z^n$, $X = \Z^n_+$, $X = [\0,\u]_\Z$, $X=[-\u,\u]_\Z$ or
  $X = [\l,\u]_\Z$ (with $\l,\u\in\Z^n$). Then, for every $\b\in\Z^m$, the
  LP relaxation $\ProbLP{1}{X}$ has an integral optimal (vertex) solution
  (in $X$).
\end{proposition}
\begin{proof}
  For $X = \Z^n$ and $X = \Z^n_+$, the results immediately follow from
  Proposition~\ref{prop:unimodularA}, since total unimodularity naturally
  implies unimodularity. The other results follow along the same lines as
  in the proof of Prop.~\ref{prop:unimodularA} by rewriting $\ProbLP{1}{X}$
  as LPs whose feasible sets are polyhedra which are integral for every
  $\b\in\Z^m$ and $\l,\u\in\Z^n$ if and only if $\A$ is totally
  unimodular. (The latter well-known characterizations can be found, e.g.,
  in \cite{Sch86}.) We omit the details to avoid repetition.
\end{proof}
\medskip

\begin{remark}
  Note that, while Propositions~\ref{prop:unimodularA} and~\ref{prop:tuA}
  do not assert \emph{unique} recoverability of $\hat{\x}\in X$ by solving
  $\ProbLP{1}{X}$ with $\b\define\A\hat{\x}$, they nevertheless guarantee
  that a feasible integral vector with the same $\ell_1$-norm as $\hat{\x}$
  can be found efficiently. Also, unlike typical recovery conditions in
  compressed sensing, the requirement of (total) unimodularity can be
  checked in polynomial time, cf.\ Seymour~\cite{Sey80},
  Truemper~\cite{Tru90} and Walter and Truemper~\cite{WalT13}.
\end{remark}

If one is interested in solution integrality of $\ProbLP{1}{X}$ for a
\emph{specific} $\b$ only, (total) unimodularity can be weakened to
requiring \emph{total dual integrality}: For $\A\in\Q^{m\times n}$,
$\b\in\Q^m$, the system $\A\x\leq\b$ is totally dual integral (TDI) if
for every $\c\in\Q^n$ such that the LP
$\min\,\{\b^\top\y\suchthat\A^\top\y=\c,~\y\geq\0\}$ is finite, it has
an integral optimal solution. Then, a well-known result (see, e.g.,
\cite[Corollary~5.14]{KorV12}) states that if $\A\x\leq\b$ is TDI and
$\b\in\Z^m$, all vertices of $\{\x\suchthat\A\x\leq\b\}$ are
integral. This and related results can be combined with the LP
relaxation $\ProbLP{1}{X}$ to obtain sufficient conditions for
integrality of optimal solutions similar to those presented in
Propositions~\ref{prop:unimodularA} and~\ref{prop:tuA}; for the sake
of brevity, we do not state this explicitly here. However, testing
whether an (in-)equality system is TDI is \NP-hard (see Ding et
al.~\cite{DinFZ08}), so these weaker conditions are harder to verify
(for a given instance).
\smallskip

Note also that the LP relaxations have (possibly after a variable
split) objective function coefficients $\c=\vones$. Thus, TDI is a
stronger requirement, since it pertains to essentially \emph{all}
$\c$, not just one specific one. Nevertheless, we can make use of
total dual integrality to obtain characterizations of relaxation
solution integrality for every $\b\in\Z^m$ by requiring the
description of the \emph{dual} polyhedron (in which $\c=\vones$ acts
as the right hand side vector) to be TDI. For the sake of exposition,
we do not go into full generality, but will consider only binary
matrices $\A$ in the remainder of this subsection.

Let us start by considering $\ProbLP{1}{\Z^n_+} =
\Prob{1}{\R^n_+}$. We need some more terminology (see~\cite{KorV12}
and Cornu\'{e}jols~\cite{Cor01} for more details): Given a (simple,
undirected) graph $G=(V,E)$, the \emph{clique-node (incidence) matrix}
$\A_G$ of $G$ has one column per node and one row per clique (i.e.,
complete subgraph) of $G$, with the $(i,j)$-entry equal to $1$ if
clique $i$ contains node $j$, and zero otherwise.  Further, recall
that a graph~$G$ is called \emph{perfect} if for every node-induced
subgraph $H$ of $G$, the chromatic number $\chi(H)$ equals the clique
number $\omega(H)$ (i.e., the minimal number of colors needed to color
the nodes of $H$ such that no neighbors have the same color coincides
with the cardinality of a maximum clique in~$H$).

\begin{proposition}\label{thm:P1N}
  Let $\A^\top\in\{0,1\}^{n\times m}$ be the clique-node matrix of a
  perfect graph. Then, for every $\b\in\Z^m$, $\ProbLP{1}{\Z^n_+}$ has
  integral optimal solutions.
\end{proposition}

For the proof, we need the following well-known result; we also provide a
proof, as it is not given in~\cite{Cor01} but useful to show the above
proposition.
\begin{lemma}[{\cite[Exercise~3.6]{Cor01}}]\label{prop:perfectTDIineq}
  Let $G=(V,E)$ be a perfect graph with clique-node incidence matrix
  $\A_G$. Then, the system $\A_G\, \y\leq\vones$, $\y\geq\0$ is TDI.
\end{lemma}
\begin{proof}
  Recall that $G$ is perfect if and only if its complement graph
  $\overline{G}$ is perfect (see, e.g., \cite[Theorem~3.4]{Cor01})
  and that cliques in $G$ correspond exactly to stable sets in
  $\overline{G}$. Hence, the system $\A_G\,\y\leq\vones$, $\y\geq\0$ is equivalent to
  \begin{equation*}
    \sum_{i\in S}y_i\leq 1\quad\forall\,\text{stable sets }S\text{ of }G,\quad\y\geq\0.
  \end{equation*}
  Suppose $\w\in\Z^V$ and consider the linear program
  \[
  \max\;\{\w\T\y \suchthat \A_G\, \y\leq\vones,\,\y\geq\0\},
  \]
  whose dual program is given by
  \begin{equation}\label{eq:perfectTDIineqLPdual}
    \min\;\Big\{\vones\T\x \suchthat \sum_{S\ni i}x_S\geq w_i\quad\forall\,i\in V,~\x\geq\0\Big\}.
  \end{equation}
  (Note that here, $x_S$ is the component of $\x$ associated with $S$.)

  We proceed to show that~\eqref{eq:perfectTDIineqLPdual} has an
  integral optimal solution for every $\w\in\Z^V$. Without loss of generality,
  we may assume that $\w\geq\vones$ (if $w_i\leq 0$, the corresponding
  constraint is automatically satisfied and can be omitted).

  Let $G'=(V',E')$ be the graph obtained from $G$ by adding $w_i-1$
  copies of each node $i\in V$ along with edges connecting each node
  copy to the respective original node and all its neighbors
  (including the other node copies). By the Replication Lemma (see,
  e.g., \cite[Lemma~3.3]{Cor01}), $G'$ is perfect. With every node
  $i\in V$, we thus associate the set $W_i$ with $\card{W_i}=w_i$ of
  nodes in $V'$; indeed, $V'=\bigcup_{i\in V}W_i$. Now, the LP
  \begin{equation}\label{eq:perfectTDIineqLPdualCopies}
    \min\;\Big\{\vones\T\x'\suchthat \sum_{S'\ni j}x'_{S'}\geq 1\quad\forall\,j\in V',~\x'\geq\0\Big\},
  \end{equation}
  where the sum is over the stable sets of $G'$, is equivalent
  to~\eqref{eq:perfectTDIineqLPdual} in the sense that feasible
  solutions of one problem can be transferred directly to feasible
  solutions of the other:
  To see this, note that we may identify a stable set $S$ in $G$ with
  all stable sets $S'$ in $G'$ whose nodes lie in $\bigcup_{i\in
    S}W_i$, and vice versa. More precisely, define 
  \[
  \rho(S)\define\{S' \text{ stable set in } G' \suchthat\card{S'}=\card{S},\,S'\cap W_i\neq\emptyset\quad\forall\,i\in S\}
  \]
  as the set of stable sets in $G'$ corresponding to a stable set $S$
  in $G$. Conversely, for any stable set $S'$ in $G'$ there exists a
  unique stable set $S$ in $G$ such that $S'\in\rho(S)$. Moreover, it holds that
  \[
  \card{\rho(S)} = \prod_{i\in S}\card{W_i}=\prod_{i\in S}w_i.
  \]

  If $\x$ is feasible for~\eqref{eq:perfectTDIineqLPdual}, then
  $\x'$ defined by $x'_{S'}\define \tfrac{1}{\card{\rho(S)}} x_S$,
  where $S'\in\rho(S)$, is feasible
  for~\eqref{eq:perfectTDIineqLPdualCopies} and has the same objective
  value. Indeed, it holds that for every $j\in W$,
  \begin{align}
    \nonumber\sum_{S'\ni j}x'_{S'}=&\sum_{S \ni i:\,j\in W_i}\; \prod_{k\in S,\,k\neq i}w_k\,\frac{x_S}{\card{\rho(S)}}=\sum_{S\ni i:\,j\in W_i}\frac{x_S}{w_i}\geq\frac{w_i}{w_i}=1\\
    \label{eq:objequal}\text{and}\quad \sum_{S'}x'_{S'}=&\sum_{S}\sum_{S'\in\rho(S)}x'_{S'}=\sum_{S}\sum_{S'\in\rho(S)}\frac{x_S}{\card{\rho(S)}}=\sum_{S}x_S.
  \end{align}
  Similarly, if $\x'$ is feasible
  for~\eqref{eq:perfectTDIineqLPdualCopies}, then $\x$ given by
  $x_S\define\sum_{S'\in\rho(S)}x'_{S'}$ is feasible
  for~\eqref{eq:perfectTDIineqLPdual} with the same objective value: Feasibility follows from
  \[
  \sum_{S\ni i}x_S=\sum_{S\ni i}\sum_{S'\in\rho(S)}x'_{S'}=\sum_{j\in W_i}\sum_{S'\ni j}x'_{S'}\geq\sum_{j\in W_i}1=w_i,
  \]
  while~\eqref{eq:objequal} shows equality of the objective values.

  Now, consider a $\chi(G')$-coloring of $G'$, i.e., a partition of
  the node set $V'$ into disjoint stable sets
  $S'_1,\dots,S'_{\chi(G')}$. Setting $x'_{S'_t}=1$ for all
  $t\in[\chi(G')]$ (and $x'_{S'}=0$ for all other $S'$) yields a
  feasible solution $\x'$
  of~\eqref{eq:perfectTDIineqLPdualCopies}. Because $G'$ is perfect,
  $\x'$ is actually optimal: Any incidence vector of a clique $C'$ in
  $G'$ is feasible for the dual program
  of~\eqref{eq:perfectTDIineqLPdualCopies} with objective value equal
  to the number of elements in $C'$. Since $\chi(G')=\omega(G')$ by
  definition of perfectness, the objective values for the coloring
  above and any maximum clique coincide. Thus, by strong duality,
  $\x'$ is optimal for~\eqref{eq:perfectTDIineqLPdualCopies} and
  consequently, so is the corresponding solution $\x$
  for~\eqref{eq:perfectTDIineqLPdual}, which concludes the proof.
\end{proof}
\medskip

\begin{proof}[Proof of Proposition~\ref{thm:P1N}]
  Let $\A\T$ be the clique-node matrix of a perfect graph. Then, the
  dual problem of $\ProbLP{1}{\Z^n_+}$ can be written as
  \[
  \max\;\big\{\b\T\y\suchthat \A\T\y\leq\vones\big\}\quad\Leftrightarrow\quad\max\;\Big\{\b\T\y\suchthat \sum_{i\in C}y_i\leq 1~\forall\,\text{cliques }C\text{ of }G\Big\}.
  \]
  Tracing the above proof of Lemma~\ref{prop:perfectTDIineq}, it is easy to
  see that the solution $\x$ for~\eqref{eq:perfectTDIineqLPdual}
  constructed there satisfies all inequality constraints of that problem
  with equality; consequently, the system $\A\T\y\leq\vones$ (without
  nonnegativity of $\y$) is also TDI.  Hence, by definition of total dual
  integrality, $\ProbLP{1}{\Z_+^n}$ has integral optimal solutions for
  every $\b\in\Z^m$.
\end{proof}
\medskip

If $\A\T$ is neither (totally) unimodular nor the clique-node matrix of
a perfect graph, integrality of the LP relaxation solutions is indeed
not ensured in general, as the following example shows.
\begin{example}\label{ex:binaryIntVSCont}
  Consider
  \[
  \A = \begin{pmatrix}
    1 & 1 & 0 \\
    0 & 1 & 1 \\
    1 & 0 & 1
    \end{pmatrix}
  \]
  and $\b = \vones \in \R^3$. In this case, $\{\x\in\Z^3_+ \suchthat \A
  \x = \b\}$ is empty, while there exists a (unique) continuous
  solution $\x = \tfrac{1}{2} \cdot \vones$. Moreover, consider
  \[
  \A = \begin{pmatrix}
    1 & 1 & 0 & 0 & 0 & 0 \\
    1 & 0 & 1 & 0 & 0 & 0 \\
    0 & 1 & 1 & 1 & 0 & 0 \\
    0 & 1 & 1 & 0 & 1 & 0 \\
    0 & 1 & 1 & 0 & 0 & 1
    \end{pmatrix}
  \]
  and $\b=\vones\in\R^6$. Here, $\x = (1, 0, 0, 1, 1, 1)\T$ is an
  optimal solution to $\Prob{1}{\Z^6_+}$. However, $\x =
  (\tfrac{1}{2}, \tfrac{1}{2}, \tfrac{1}{2}, 0, 0, 0)\T$ is an optimal
  solution to $\Prob{1}{\R^6_+}$ (with fewer nonzeros). 
\end{example}

\begin{remark}
  Note that in Proposition~\ref{thm:P1N}, $\A\T$ is the clique-node matrix
  with respect to \emph{all} cliques in a perfect graph. This differs from
  the usual theory, which allows for restricting to (inclusion-wise)
  maximal cliques (see the already accordingly restricted definition of
  clique-node matrices in~\cite{Cor01}). Indeed,
  Proposition~\ref{prop:perfectTDIineq} remains true under such a
  restriction, but this does not carry over to Proposition~\ref{thm:P1N}.
  Also, one can easily find examples which show that one does not
  necessarily need to include all cliques to achieve total dual integrality
  of $\A^\top \y\leq\vones$, which in turn (with binary $\A$) does not
  imply that $\A^\top$ is the clique-node matrix of a perfect graph.
\end{remark}
  
We can extend the results from Proposition~\ref{thm:P1N} to a sufficient
condition for solution integrality for $\ProbLP{1}{\Z^n}=\Prob{1}{\R^n}$.  

\begin{proposition}\label{thm:P1Z}
  Let $\A^\top\in\{0,1\}^{n\times m}$ be the clique-node matrix of a
  perfect graph. Then, for every $\b\in\Z^m$, $\ProbLP{1}{\Z^n}$ has
  integral optimal solutions.
\end{proposition}
\begin{proof}
  By means of a standard variable split $\x=\x^+-\x^-$ with
  $\x^\pm\define\max\,\{\0,\pm\x\}$ (component-wise), we can rewrite
  $\ProbLP{1}{\Z^n}$ as the LP
  \begin{equation}\label{eq:Prob1Rsplit}
    \min\;\big\{\vones^\top \x^++\vones^\top \x^-\suchthat\A\x^+-\A\x^-=\b,\,\x^+\geq\0,\,\x^-\geq\0\big\}.
  \end{equation}
  Without loss of generality, we may assume that
  $\b = ((\b^+)\T, (\b^-)\T)\T$ with
  $\b^+\geq\0$, $\b^-\leq\0$ (permuting rows, if necessary). Let
  $\B=(\A,-\A)$, and denote by $\B^+$ and $\B^-$ the submatrices
  corresponding to the rows associated with $\b^+$ and $\b^-$,
  respectively. Furthermore, we write $\B^+=(\B^+_+,\B^+_-)$ and
  $\B^-=(\B^-_+,\B^-_-)$ to distinguish the respective columns
  corresponding to $\x^+$ and $\x^-$. We can now
  rewrite~\eqref{eq:Prob1Rsplit} and relax its constraints as follows:
  \begin{align*}
    &\min\;\big\{\vones^\top \x^++\vones^\top \x^-\suchthat\A\x^+-\A\x^-=\b,\,\x^+\geq\0,\,\x^-\geq\0\big\}\\
    = &\min\;\left\{\vones^\top\x^+ +\vones^\top\x^-\suchthat\left(\hspace*{-0.4em}\begin{array}{rr}\B^+_+ & \hspace*{-0.4em}\B^+_-\\\B^-_+ & \hspace*{-0.4em}\B^-_-\end{array}\hspace*{-0.4em}\right)\left(\hspace*{-0.4em}\begin{array}{r}\x^+\\\x^-\end{array}\hspace*{-0.4em}\right)=\left(\hspace*{-0.4em}\begin{array}{r}\b^+\\\b^-\end{array}\hspace*{-0.4em}\right),\,\x^+\geq\0,\,\x^-\geq\0\right\}\\
    \geq &\min\;\big\{\vones^\top\x^+\suchthat \B^+_+\, \x^+ + \B^+_-\,\x^- =\b^+,\,\x^+\geq\0,\,\x^-\geq\0\big\}\\
    &\quad +\min\;\big\{\vones^\top\x^-\suchthat \B^-_+\, \x^+ + \B^-_-\, \x^- =\b^-,\,\x^+\geq\0,\,\x^-\geq\0\big\}\\
    \geq &\min\;\big\{\vones^\top\x^+\suchthat \B^+_+\, \x^ + \geq \b^+ - \B^+_-\, \x^-,\,\x^+\geq\0,\,\x^-\geq\0\big\}\\
    &\quad +\min\;\big\{\vones^\top\x^-\suchthat \B^-_+\, \x^+ \leq \b^- - \B^-_-\, \x^-,\,\x^+\geq\0,\,\x^-\geq\0\big\}.
  \end{align*}
  Observing that $\B^+_-\,\x^-\leq\0\leq\B^-_+\,\x^+$ (since $\A$ is
  binary), we can further relax the last two programs as
  \begin{align*}
    &\min\;\big\{\vones^\top\x^+\suchthat \B^+_+\,\x^ + \geq \b^+ - \B^+_-\,\x^-,\,\x^+\geq\0,\,\x^-\geq\0\big\}\\
    \geq\, &\min\;\big\{\vones^\top\x^+\suchthat \B^+_+\, \x^ + \geq \b^+,\,\x^+\geq\0\big\}
  \end{align*}
  and
  \begin{align*}
    &\min\;\big\{\vones^\top\x^-\suchthat \B^-_+\, \x^ + \leq \b^- - \B^-_-\, \x^-,\,\x^+\geq\0,\,\x^-\geq\0\big\}\\
    \geq\, &\min\;\big\{\vones^\top\x^-\suchthat -\B^-_-\, \x^- \geq -\b^-,\,\x^-\geq\0\big\}.
  \end{align*}
  Thus, the optimal objective function value of \eqref{eq:Prob1Rsplit} is
  bounded from below by the sum of the optimal values of two linear
  programs, each of which can easily be seen to be associated with a
  (generalized) set covering problem.

  By definition, $(\B^+_+)^\top$ is a matrix whose rows are the
  incidence vectors of all cliques of a perfect graph (a node-induced
  subgraph of the graph represented by $\A^\top$); the same holds for
  $(-\B^-_-)^\top$. Hence, by Lemma~\ref{prop:perfectTDIineq}
  and the proof of Proposition~\ref{thm:P1N}, both LPs have optimal
  integral solutions---say, $\x_*^+$ and $\x_*^-$---that satisfy all
  inequality constraints with equality. 

  Moreover, for every column of $\A$ representing a clique (and thus
  for any column of $\B^+_+$ or $\B^-_-$ representing a clique), there
  is another column for every subclique. Therefore, the solutions
  $\x_*^+$ and $\x_*^-$ may be chosen such that
  $\B^-_+\, \x_*^+ = \0 = \B^+_-\, \x_*^-$.

  It remains to observe that $\x_*\define\x_*^+-\x_*^-$ is an integral
  feasible solution for $\ProbLP{1}{\Z^n}$, and since it achieves the
  lower bound given by the two LPs derived above, it is, in fact,
  optimal.
\end{proof}

\end{appendices}

\end{document}